\documentclass[sigplan,10pt]{acmart}

\acmYear{2026}\copyrightyear{2026}
\setcopyright{cc}
\setcctype[4.0]{by}
\acmConference[EUROSYS '26]{European Conference on Computer Systems}{April 27--30, 2026}{Edinburgh, Scotland Uk}
\acmBooktitle{European Conference on Computer Systems (EUROSYS '26), April 27--30, 2026, Edinburgh, Scotland Uk}
\acmDOI{10.1145/3767295.3803571}
\acmISBN{979-8-4007-2212-7/26/04}

\usepackage{booktabs}
\usepackage{tikz}
\usepackage{comment}
\usepackage{amsmath}
\usepackage{caption}
\usepackage{subcaption}
\usepackage{graphicx}
\usepackage{float}
\usepackage{algorithm}
\usepackage{algorithmicx}
\usepackage[noend]{algpseudocode}
\algrenewcommand\textproc{}
\algnewcommand{\LineComment}[1]{\State \(//\) #1}
\usepackage[normalem]{ulem}
\usepackage{bbding}
\usepackage{todonotes}
\usepackage{filecontents}
\usepackage{minted}
\usepackage{enumitem}
\usepackage[bottom]{footmisc}
\usepackage{hyperref}
\usepackage{cleveref}
\usepackage{pifont}
\usepackage{multirow}
\usepackage[dvipsnames,svgnames]{xcolor}
\usepackage{outlines}
\usepackage{makecell}
\usepackage{wasysym}
\usepackage[framemethod=TikZ]{mdframed}

\usepackage{listings}
\usepackage{xcolor}

\newcolumntype{C}[1]{>{\centering\arraybackslash}m{#1}}

\newcommand{\sysname}{Arena}

\makeatletter
\renewcommand{\paragraph}{%
  \@startsection{paragraph}{4}%
  {\z@}{1ex \@plus 1ex \@minus 1ex}{-1em}%
  {\normalfont\normalsize\bfseries}%
}

\begin{document}

\title[]{\sysname{}: Efficiently Training Large Models via Dynamic Scheduling and Adaptive Parallelism Co-Design}

\author{Chunyu Xue}
\email{dicardo@sjtu.edu.cn}
\affiliation{%
  \institution{Shanghai Jiao Tong University}
  \country{}
}

\author{Weihao Cui}
\email{weihao@sjtu.edu.cn}
\affiliation{%
  \institution{Shanghai Jiao Tong University}
  \country{}
}

\author{Quan Chen}
\email{chen-quan@cs.sjtu.edu.cn}
\affiliation{%
  \institution{Shanghai Jiao Tong University}
  \country{}
}

\author{Chen Chen}
\email{chen-chen@sjtu.edu.cn}
\affiliation{%
  \institution{Shanghai Jiao Tong University}
  \country{}
}

\author{Han Zhao}
\email{zhaohan_miven@sjtu.edu.cn}
\affiliation{%
  \institution{Shanghai Jiao Tong University}
  \country{}
}

\author{Shulai Zhang}
\email{zslzsl1998@sjtu.edu.cn}
\affiliation{%
  \institution{Shanghai Jiao Tong University}
  \country{}
}

\author{Linmei Wang}
\email{wanglm22@lenovo.com}
\affiliation{%
  \institution{Lenovo Research}
  \country{}
}

\author{Yan Li}
\email{yanli5@microsoft.com}
\affiliation{%
  \institution{Microsoft}
  \country{}
}

\author{Limin Xiao}
\email{xiaolm@lenovo.com}
\affiliation{%
  \institution{Lenovo Research}
  \country{}
}

\author{Weifeng Zhang}
\email{weifengz@lenovo.com}
\affiliation{%
  \institution{Lenovo Research}
  \country{}
}

\author{Jing Yang}
\email{jyang23@gzu.edu.cn}
\affiliation{%
  \institution{Guizhou University}
  \country{}
}

\author{Bingsheng He}
\email{hebs@comp.nus.edu.sg}
\affiliation{%
  \institution{National University of Singapore}
  \country{}
}

\author{Minyi Guo}
\email{guo-my@cs.sjtu.edu.cn}
\affiliation{%
  \institution{Shanghai Jiao Tong University}
  \country{}
}

\renewcommand{\shortauthors}{Xue et al.}

\renewcommand{\shorttitle}{\sysname{}: Training Large Models via Dynamic Scheduling and Adaptive Parallelism Co-Design}

\begin{abstract}

Efficiently training large-scale models (LMs) in GPU clusters involves two separate avenues: inter-job dynamic scheduling and intra-job adaptive parallelism (AP).
However, existing dynamic schedulers struggle with large-model scheduling due to the mismatch between static parallelism (SP)-aware scheduling and AP-based execution, leading to cluster inefficiencies such as degraded throughput and prolonged job queuing.
This paper presents \sysname{}, a large-model training system that co-designs dynamic scheduling and adaptive parallelism to achieve high cluster efficiency. 
To reduce scheduling costs while improving decision quality, \sysname{} designs low-cost, disaggregated profiling and AP-tailored, load-aware performance estimation, while unifying them by sharding the joint scheduling-parallelism optimization space via a grid abstraction.
Building on this, \sysname{} dynamically schedules profiled jobs in elasticity and heterogeneity dimensions, and executes them using efficient AP with pruned search space.
Evaluated on heterogeneous testbeds and production workloads, \sysname{} reduces job completion time by up to $49.3\%$ and improves cluster throughput by up to $1.60\times$.


\end{abstract}



\maketitle 

\section{Introduction}\label{sec:intro}

Training jobs of large-scale models (LMs), such as GPT~\cite{gpt-3} and MoE~\cite{gshard}, are vital workloads in production GPU clusters.
With the rapid evolution of hardware architecture (e.g., NVIDIA Blackwell~\cite{Blackwell}), these clusters continually integrate new-generation GPUs with diverse compute capabilities and memory sizes, resulting in expanded cluster scale and heterogeneous resource types~\cite{pai,helios,heteroscale}.
For instance, a production cluster~\cite{bootseer} has thousands of GPUs with a mixture of NVIDIA H800~\cite{Hopper}, A100~\cite{Ampere}, and V100 GPUs~\cite{Volta}.

To efficiently exploit these cluster resources, instead of user-specified rigid allocation~\cite{jeon2019analysis}, recent schedulers such as ElasticFlow~\cite{elasticflow} and Sia~\cite{sia} \textit{dynamically} (re)assign them to concurrent jobs, by adjusting the number (``elasticity'') and type (``heterogeneity'') of GPUs~\cite{gandiva,elasticflow,gavel,sia,pollux,easyscale,lucid,gandiva_fair}.
From the job perspective, multiple parallelism dimensions, including data~\cite{li2014scaling}, model~\cite{megatron,zero}, and pipeline~\cite{pipedream} parallelism, are used to execute LMs in a hybrid manner.
Frameworks like Alpa~\cite{alpa} automatically explore the optimal hybrid parallelism plan under specific model and resource configurations~\cite{alpa,nnscaler,metis}, referred to as \textit{adaptive parallelism} (AP) in this paper.
Given the fixed model and resource allocation, it formulates a search space of parallelism plans and identifies the optimal plan via costly profiling~\cite{alpa}.

\begin{figure}
\centering
\includegraphics[width=.98\linewidth]{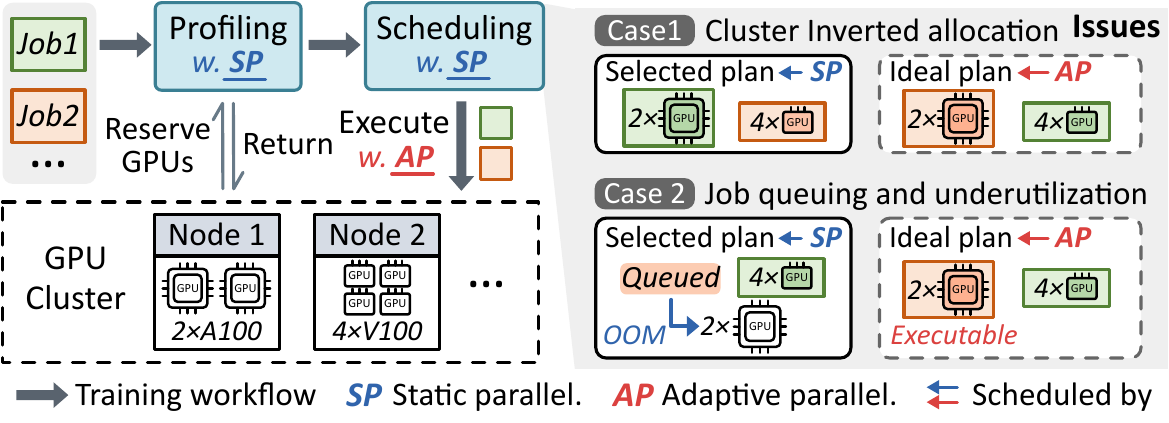}
\vspace{-2mm}
\caption{Workflow of dynamic scheduling based on static parallelism (SP) performance while executing with adaptive parallelism (AP). 
Our work targets optimizing this workflow.
}
\label{fig:dp_issues}
\vspace{-4mm}
\end{figure}

However, prior dynamic schedulers~\cite{elasticflow,sia,pollux,gavel,easyscale} disregard adaptive parallelism when allocating resources, causing reduced cluster efficiency, as depicted in Figure~\ref{fig:dp_issues}.
For training jobs, the scheduler allocates resources based on the performance of statically decided parallelism plans (or \textit{static parallelism}, SP), typically data parallelism (DP), which is attained from model \textit{profiling}~\cite{elasticflow,lucid} or performance \textit{estimation}~\cite{sia,gavel}.
For example, 
ElasticFlow profiles jobs with DP across allocable resources in 10 minutes, 
Sia estimates multi-GPU DP throughput by linearly scaling single-GPU one with the GPU count.
Given the dynamicity of AP (\S\ref{sec:contradiction}), such \textit{mismatch between SP-aware scheduling and AP-based execution} leads to significant performance acquisition errors and degrades cluster efficiency.
We empirically demonstrate this through two cases.
First, the ideal resource allocation in the cluster could be ``inverted'' by the performance discrepancy between SP and AP execution (\texttt{Case1}). 
Second, the resource demands of LM jobs are overestimated (\texttt{Case2}), as static DP consumes the most memory among all parallelism.
This results in the ideal plan being omitted, bringing more job queuing and resource underutilization in clusters.

Co-designing dynamic scheduling and adaptive parallelism is essential to mitigate these inefficiencies, though we find it challenging and lacking prior work.
In this context, the scheduling and parallelism spaces form a joint optimization space via outer product (\S\ref{sec:grid}).
A strawman approach is exhaustively profiling jobs with AP across allocable resources before scheduling, as in ElasticFlow and Lucid~\cite{lucid}.
While intuitive, this incurs excessive time and hardware costs, as profiling needs reserving considerable GPUs for the time-consuming AP search (e.g., 20 minutes per allocable resource~\cite{alpa}). 
Given the prevalent GPU scarcity in production clusters~\cite{pai,hived}, this approach is impractical due to prolonged job queuing and resource contention.
Another thought is analytically estimating job AP performance before scheduling, and executing jobs with AP at runtime, as in Sia and Gavel~\cite{gavel}. 
The execution is online profiled to refine estimates for better subsequent rescheduling.
Despite its low cost, accurate estimation is non-trivial due to the shifting parallelism plan and scaling diminishing returns under dynamic resources.
Inaccurate estimation undermines scheduling quality and cluster efficiency, which cannot be refined via frequent job rescheduling due to the high cost of AP search.

Our key insight to improve cluster efficiency and reduce scheduling costs is that, the scheduler can hierarchically unify low-cost profiling and AP-tailored estimation to efficiently navigate the joint scheduling-parallelism space.
It is based on our observation that, given a model with specific resource, the performance relationship between two parallelism plans in AP search space can be analytically discerned without precise latencies, when the number of stages is fixed (\S\ref{sec:grid}).
Building on this insight, we present \textbf{\sysname{}}, a co-designed training system that dynamically schedules and efficiently executes large models with adaptive parallelism.
Inspired by grid sampling~\cite{grid_sample}, \sysname{} shards the joint space into subspaces (``grids'').
A grid of a job includes scheduling-parallelism plans with identical resource and \textit{pipeline degree} (the number of stages).
Within a grid, a near-optimal parallelism plan (``proxy plan'') is analytically estimated; among grids, proxy plans are profiled and used for scheduling. 

Concretely, for each grid, \sysname{} identifies its proxy plan via roofline-based plan generation and Pareto frontier deduction in execution-free manner (\S\ref{sec:estimation}).
Given proxy plans, \sysname{} profiles each via operator disaggregation and single-device profiling on fragmented resources (\S\ref{sec:profiling}). 
With profiled jobs, \sysname{} dynamically schedules them in elasticity and heterogeneity dimensions via job launching and scaling mechanisms, and a generalized event-driven policy with diverse objectives (\S\ref{sec:sched}).
After scheduled, \sysname{} executes jobs with efficient AP pruned by estimation and profiling results (\S\ref{sec:runtime}).
The main contributions of this work are:

\begin{itemize}[itemsep=0pt, parsep=0pt, labelsep=5pt, leftmargin=*, topsep=3pt,partopsep=0pt]
\itemsep3pt

    \item We reveal the issues of existing SP-aware schedulers in large-model scheduling, and identify the opportunity and challenges of joint scheduling-parallelism optimization.
	
    \item We propose \sysname{}, a multi-model training system for large models, which co-designs dynamic scheduling and adaptive parallelism to improve cluster efficiency.

    \item We design low-cost, disaggregated profiling and AP-tailored, load-aware estimation, unified by a novel grid abstraction for efficient and precise job performance acquisition.
    
    
\end{itemize}

We implement \sysname{} and evaluate it with various large models and production traces, in two real-world heterogeneous clusters with 64 and 384 GPUs of 2 types and a simulated cluster with 1,280 GPUs of 4 types. 
Extensive experiments show that \sysname{} achieves up to $1.60\times$ higher cluster throughput, reduces queuing delay by $74.9\%$ and job completion time (JCT) by $49.3\%$, compared to four baselines.
\sysname{} is open-sourced at \url{https://github.com/sjtu-epcc/arena}.

\section{Background and Motivation}\label{sec:moti}

\subsection{Training Large Models in Clusters}\label{sec:background}


\paragraph{Parallelism Strategies.}
Various parallelism strategies are proposed for multi-GPU model training.
Data parallelism (DP)~\cite{li2014scaling} splits batches across model replicas for concurrent computation, yet consumes substantial memory.
Pipeline parallelism (PP)~\cite{gpipe,pipedream,terapipe} groups operators into stages and splits a batch into microbatches, pipelining execution to improve throughput, but causes device idling.
Model parallelism (MP), including tensor parallelism~\cite{megatron} and ZeRO~\cite{zero,zero_offload}, splits operators or intermediate states across GPUs to reduce memory, at the cost of considerable communication.

Given the above pros and cons, the optimal parallelism plan is a hybrid combination that hinges on models, inputs, and hardware.
To identify it, adaptive parallelism (AP) explores the search space of parallelism plans based on extensive profiling, by partitioning the model into pipeline stages, and parallelizing each across their assigned GPUs~\cite{alpa,nnscaler}.
Other strategies such as sequence parallelism~\cite{seq_parallelism} and expert~ parallelism~\cite{tutel} follow the paradigm of MP by splitting tensors across GPUs with communication incurred~\cite{metis,nnscaler}.

\begin{figure}
\centering
\includegraphics[width=\linewidth]{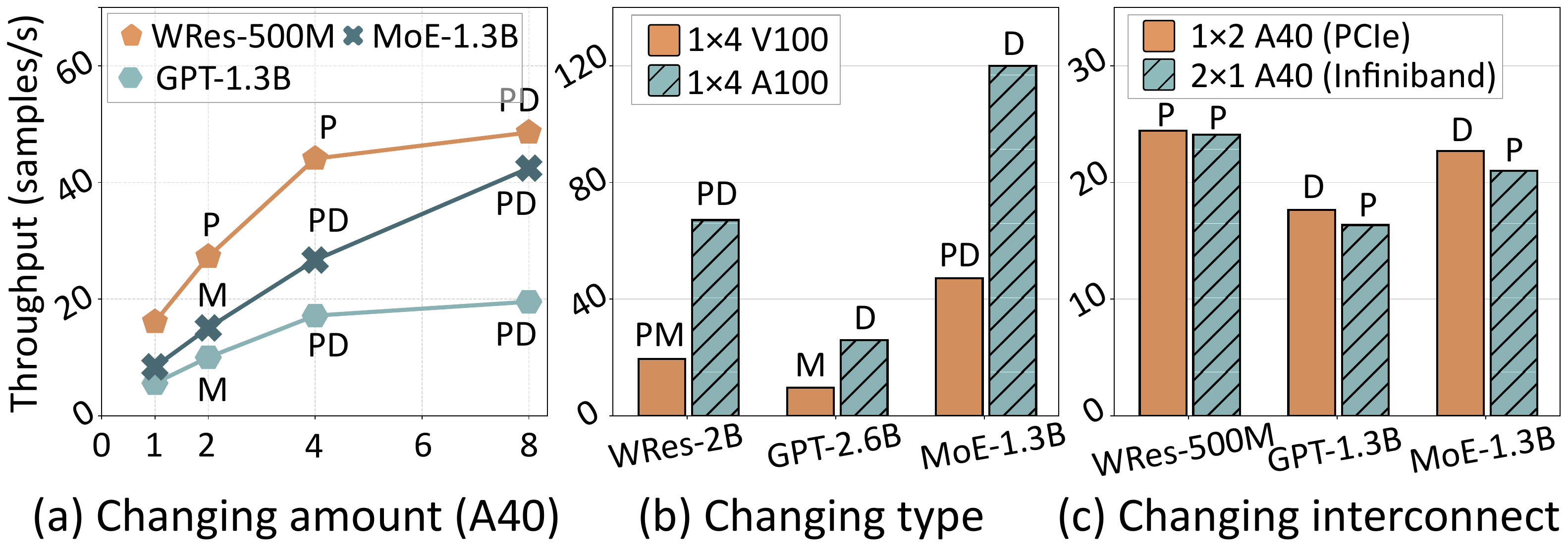}
\vspace{-6mm}
\caption{Benchmarking adaptive parallelism across various models and hardware. \texttt{P}/\texttt{D}/\texttt{M} denote pipeline, data and model parallelism. The optimal parallelism plan of AP is annotated.}
\label{fig:scaling}
\vspace{-2mm}
\end{figure}

\paragraph{Cluster Scheduling.}
Traditionally, resources are rigidly allocated on-demand, causing head-of-line blocking and resource fragmentation~\cite{pai,easyscale}. 
To mitigate the issues in homogeneous clusters, recent schedulers~\cite{elasticflow,lucid} dynamically adjust the number of GPUs for training jobs (\textbf{``elasticity''}).

As GPU manufacturers continue developing and rolling out new-generation GPUs (e.g., NVIDIA Hopper~\cite{Hopper}, Blackwell~\cite{Blackwell}), many production clusters continually integrate heterogeneous GPUs~\cite{bootseer,helios,pai,heteroscale}.
Thus, both academic researches~\cite{gavel,sia,gandiva_fair} and industrial practice~\cite{heteroscale,bootseer} put efforts on efficiently utilizing heterogeneous hardware, including dynamically adjusting GPU types for training jobs (\textbf{``heterogeneity''}).
For instance, the clusters in~\cite{bootseer,heteroscale} consist of multiple regions, each for a job queue with a specific GPU type. 
Each cluster integrates a scheduler that allows users to specify ``any-type'' GPUs during job submission, thereby reducing job queuing and improving resource utilization.
Recent schedulers~\cite{sia,easyscale} combine elasticity and heterogeneity dimensions for more flexible scheduling.

\subsection{Problems in Scheduling Large Models}\label{sec:contradiction}



\paragraph{SP-aware scheduling in practice.}
Current schedulers, such as ElasticFlow~\cite{elasticflow}, Sia~\cite{sia}, and EasyScale~\cite{easyscale}, disregard adaptive parallelism (AP) when allocating resources.
As common practice, they assume that all training jobs are executed with static parallelism (typically DP\footnote{Some schedulers like Sia~\cite{sia} additionally support a fixed pipeline with manually partitioned stages instead of fully flexible AP.}).
This is reasonable in scheduling conventional models (e.g., ResNet, VGG), as they perform well with DP and rarely need AP.
Another benefit is that the scheduler can acquire job performance via intuitive profiling~\cite{elasticflow,gandiva,lucid} or estimation~\cite{gavel,pollux,sia}.


\begin{figure}
\centering
\includegraphics[width=.98\linewidth]{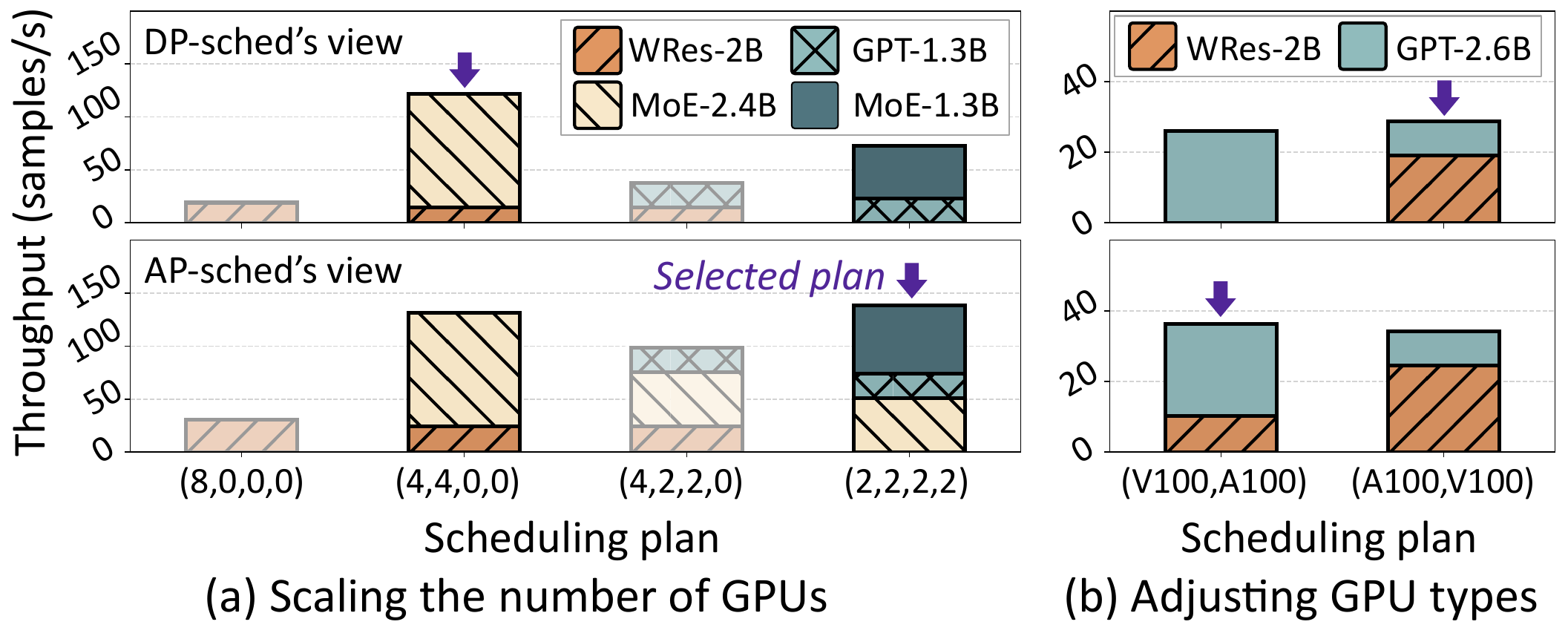}
\vspace{-2mm}
\caption{Case study of scheduling plan selection based on static data and adaptive parallelism. Missing bars indicate OOM issues. (a) \texttt{(a,b,c,d)}: \texttt{a}$\times$A100 GPUs for WRes-2B, \texttt{b}$\times$ for MoE-2.4B, \texttt{c}$\times$ for GPT-1.3B, \texttt{d}$\times$ for MoE-1.3B. (b) \texttt{(A,B)}: $4\times\texttt{A}$ GPUs for WRes-2B and $4\times\texttt{B}$ for GPT-2.6B.}
\label{fig:cluster_dp_ap}
\vspace{-2mm}
\end{figure}

\paragraph{Mismatch between SP-aware scheduling and AP-based execution.}
However, the assumed static parallelism (SP) is mismatched with the actually used parallelism when LM jobs are executed with adaptive parallelism (AP), resulting in degraded cluster efficiency.
This inefficiency primarily stems from the \textit{performance discrepancies between SP and AP execution}.
Figure~\ref{fig:scaling} presents the AP performance of various models and hardware (configured as Table~\ref{tab:sim_cluster}).
We observe that instead of fixed parallelism patterns, AP exhibits significant dynamicity across different hardware. 
For instance, with sufficient resources, DP is preferred to improve computing concurrency (e.g., 8-GPU in Figure~\ref{fig:scaling}(a)); 
with limited memory, MP is used to reduce per-GPU memory footprint (e.g., GPT-2.6B in Figure~\ref{fig:scaling}(b)); 
with limited bandwidth, PP is selected for its lowest communication (e.g., MoE-1.3B in Figure~\ref{fig:scaling}(c)).
Different models also exhibit varying parallelism preference on the same hardware.
Consequently, determining the optimal plan in practice is a complex task.

As demonstrated, the optimal parallelism plan of AP and the throughput shift significantly across different configurations, rather than remaining static.
In this case, assuming static parallelism such as DP covers only a marginal fraction of the vast search space defined by AP, in which the optimal plan hinges on multiple factors such as models, inputs, and hardware.
For this reason, SP-aware scheduling leads to performance acquisition errors when LMs are executed with AP.
Figure~\ref{fig:cluster_dp_ap} illustrates the impact of this mismatch on cluster efficiency by two frequently occurred cases, which is a quantitative case study of large-scale scheduling (\S\ref{sec:sim}).

\textbf{\textit{Case\#1}. Cluster-level inverted allocation.}
In Figure~\ref{fig:cluster_dp_ap}(a), with static DP, the throughput of plan \texttt{(4,4,0,0)} is higher than plan \texttt{(2,2,2,2)}, yet the latter is optimal with AP.
Similar observations exist between plan \texttt{(A100,V100)} and plan \texttt{(V100,A100)} in Figure~\ref{fig:cluster_dp_ap}(b).
Consequently, inverted allocation results in reduced overall throughput (e.g., $1.1\times$).

\textbf{\textit{Case\#2}. Job-level prolonged queuing and underutilization.}
Since DP consumes the most memory among all parallelism, MoE-2.4B is assigned 4 GPUs though trainable on 2 GPUs with AP (\texttt{(4,2,2,0)} in Figure~\ref{fig:cluster_dp_ap}(a)).
This omits scheduling plans with ``dense'' allocation (high concurrency), causing prolonged job queuing and resource fragmentation.
Moreover, overly allocating GPUs leads to underutilization and degraded throughput due to diminishing returns from excess resources~\cite{hardware_scaling,elasticflow}.
In Figure~\ref{fig:scaling}(a), the throughput of GPT-1.3B scales sub-linearly with the GPU count.
Given 4 GPUs, using all for GPT-1.3B yields $1.5\times$ lower throughput than using 2 GPUs each for GPT-1.3B and MoE-1.3B. 

\begin{figure}
\centering
\includegraphics[width=0.9\linewidth]{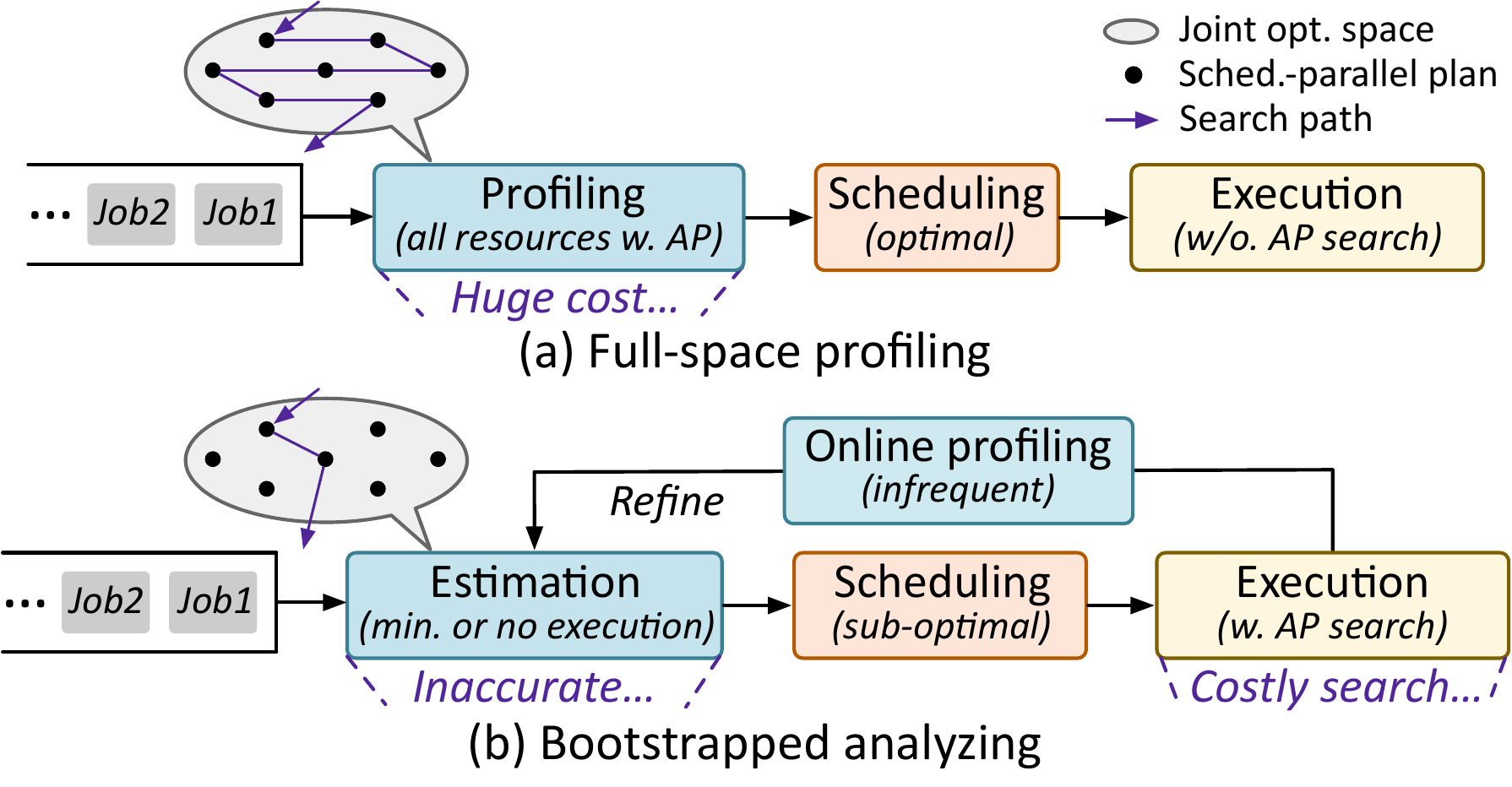}
\vspace{-2mm}
\caption{Workflows of two strawman approaches.}
\label{fig:strawman}
\vspace{-2mm}
\end{figure}

\subsection{Deficiencies of Strawman Approaches}\label{sec:strawman}

To improve cluster efficiency, the scheduler should consider adaptive parallelism (AP) performance for each job when allocating resources.
In this context, the scheduling and parallelism spaces form a joint optimization space via outer product (\S\ref{sec:grid}).
Given the limited prior work on this research topic, we in-depth analyze existing schedulers and identify two representative workflows: \textit{full-space profiling} (e.g., ElasticFlow~\cite{elasticflow}, Lucid~\cite{lucid}) and \textit{bootstrapped analyzing} (e.g., Sia~\cite{sia}, Gavel~\cite{gavel}), as depicted in Figure~\ref{fig:strawman}.
We use them as intuitive attempts to co-design scheduling and parallelism.

\paragraph{Full-space profiling results in huge time and hardware costs.} 
The full-space profiling approach traverses the joint space before scheduling, i.e., profiling jobs with AP across all allocable resources (Figure~\ref{fig:strawman}(a)).
At runtime, jobs are executed with the identified optimal plan.
This ensures decision optimality and enables rescheduling without extra AP search, yet incurring excessive time and hardware costs beforehand.

Assume each job can be assigned up to $N$ GPUs per type among $M$ GPU types.
Before scheduling, these $NM$ GPUs should be reserved for model profiling.
Let $T_{ap}$ denotes the AP search cost (e.g., 20 minutes~\cite{alpa}), the profiling cost per job is $(N + (N-1) + ... + 1) M T_{ap} = O(N^2MT_{ap})$ GPU hours~\cite{scao2022language}.
When $K$ jobs arrive, say 8 from our trace statistics~\cite{jeon2019analysis,pai,helios}, with $N=16$ and $M=4$, the scheduler requires at least $1,500$ GPU hours for profiling.

Such substantial cost renders full-space profiling impractical, making profiling cost reduction a key prerequisite in scheduling–parallelism co-design.
Specifically, available GPUs are often scarce and fragmented in real-world clusters due to the prevalence of pending jobs with large resource demands~\cite{skypilot,easyscale,pai,hived}. 
Jobs undergo much prolonged queuing to be allocated their maximal allocable resources (i.e., $NM$ GPUs) for profiling.
Moreover, substantial economic costs are incurred in this non-training phase (e.g., 5 dollars per GPU hour for AWS \texttt{pxde.24xlarge} with A100 GPUs~\cite{aws_p4}), while inter-job resource contention is exacerbated.


\paragraph{Bootstrapped analyzing results in inaccurate estimation.} 
The bootstrapped analyzing approach estimates AP performance across allocable resources before scheduling (Figure~\ref{fig:strawman}(b)).
At runtime, jobs are executed with AP and online profiled to refine estimates for better subsequent rescheduling.
There are two major types of estimation.
The first type analytically models the latency of computation operators by ``FLOPs $/$ $\{$GPU peak performance $\times$ compute utilization$\}$''~\cite{paleo,madmax}.
The utilization is empirically set by hardware profiling (e.g., $70\%$ in~\cite{madmax}).
This estimation becomes inaccurate under dynamic resources and adaptive parallelism due to the shifting compute and bandwidth utilization~\cite{neusight,centimani,alpa}. 
To illustrate this, we evaluate it using GPT-2.6B on 4 A40 GPUs, observing a $61.3\%$ estimation error with $63.9\%$ in compute (aligned with~\cite{habitat}) and $36.1\%$ in communication latency (``volume $/$ bandwidth'').

The second type is partially profiling some configurations while estimating others (e.g., Sia~\cite{sia}, Centimani~\cite{centimani}).
For instance, Sia assumes ``throughput of 2-way DP is twice that of 1-way DP'', profiling only 1-GPU throughput $\texttt{thr}_1$ per GPU type, and linearly estimating $n$-GPU throughput by $\texttt{thr}_n = \texttt{thr}_1 \times N_{gpu}$.
It becomes inaccurate due to the overlook of parallelism shifting in adaptive parallelism and diminishing returns in resource scaling~\cite{hardware_scaling}.
To illustrate this, we profile the AP throughput of GPT-1.3B from 1 to 16 GPUs, observing that the linear estimation yields errors from $1.14\times$ (2 GPUs) to $2.12\times$ (16 GPUs), with the optimal parallelism plan shifted from MP to DP and PP.




To investigate the impact of inaccurate estimation on scheduling quality, we add a knob $\eta$ to regulate the fraction of precise profiling data in the linear estimation of Sia. 
For instance, when $\eta=2$, we precisely profile 1 and 2-GPU AP throughput ($\leq 2^{\eta-1}$). 
As $\eta$ increases from 1 (original estimation) to 5 (fully precise data), the overall throughput improves by $1.19\times$, as configured in \S\ref{sec:sim}.
However, the high cost of AP search hinders frequent correction of inaccurate  estimation, ultimately degrading cluster efficiency.


\subsection{Our Approach and Challenges}\label{sec:challenges}

Analyzing the strawman approaches, we identify the \textit{contradiction} in AP-aware scheduling: performant decisions require costly profiling, whereas swift estimation leads to poor scheduling quality.
Prior works have focused on customizing scheduling algorithms, which cannot fundamentally resolve this contradiction~\cite{sia,gavel,elasticflow}.
In contrast, this work focuses on addressing this by designing principled approaches to enable efficient scheduling-parallelism co-design.


To reduce scheduling costs and ensure quality, we propose a two-fold approach that leverages the strengths of two strawmans to balance the ``cost-accuracy'' tradeoff, i.e., \textit{unifying lightweight estimation and precise profiling}.
Specifically, we find that estimation remains accurate in this context: for a model with specific resource, the relative performance between two parallelism plans in AP search space can be analytically discerned if sharing identical pipeline degree (\S\ref{sec:grid}).
Beyond that, comparing scheduling-parallelism plans with varied pipeline degrees, resources, or models (inter-job) is essential for scheduling and requires profiling to obtain precise latency.
There are three main challenges to develop and integrate this two-fold approach into cluster scheduling for efficient scheduling-parallelism co-design:
\textbf{\textit{C\#1.} Lack precise parallelism estimation tailored for AP.}
Coarse-grained methods fail to precisely estimate AP performance, as identifying optimal plans while adhering to AP optimality is non-trivial (\S\ref{sec:strawman}).
Instead of simplistic assumptions, a tailored, execution-free approach based on the above observation is needed to assess parallelism plans.

\textbf{\textit{C\#2.} Lack practical profiling with minimal time and hardware costs.}
Model-granularity profiling incurs high costs, prolonged queuing, and resource contention when AP is enabled (\S\ref{sec:strawman}).
Enhancing its production practicality requires reducing the number of occupied GPUs and the elapsed time while maintaining measurement accuracy.

\textbf{\textit{C\#3.} Efficient scheduling and low-cost AP execution.}
Given considerable number of jobs and heterogeneous resources, it is non-trivial to generate efficient scheduling plans with negligible overhead and high scalability.
For fast deployment, it is essential to reduce the search cost of AP.

\section{\sysname{} Design}\label{sec:design}


\subsection{Overview}\label{sec:overview}

\begin{figure}
\centering
\includegraphics[width=\linewidth]{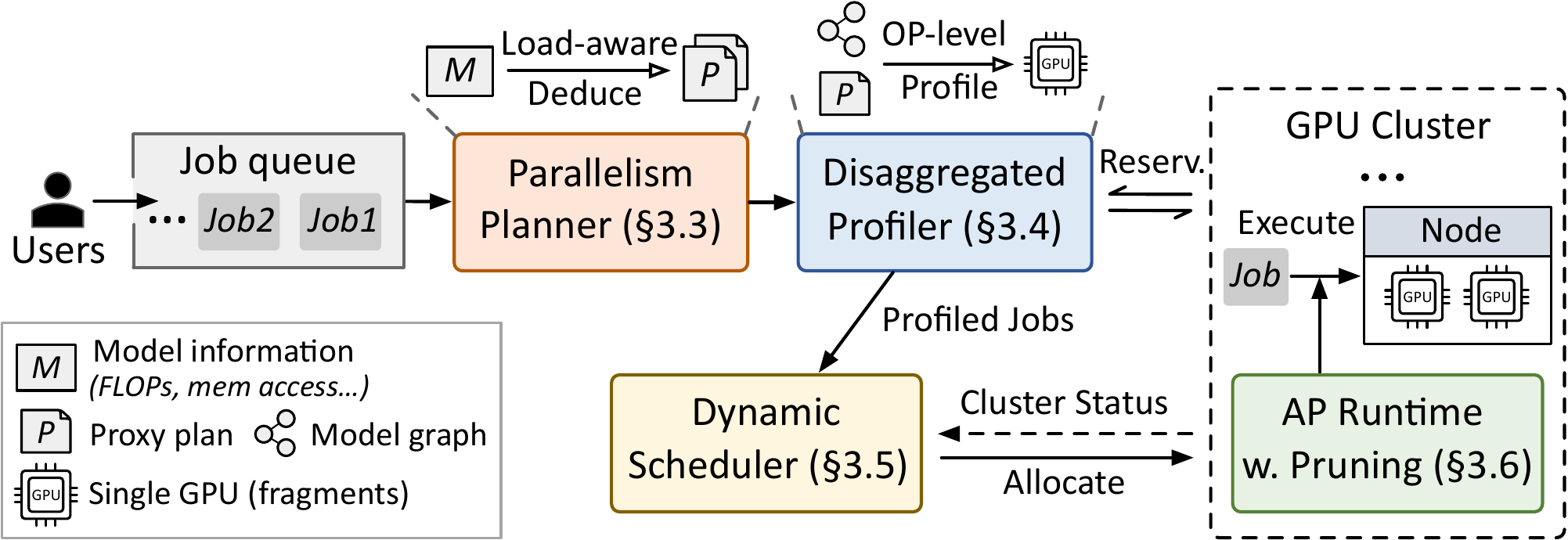}
\vspace{-6mm}
\caption{The overview of \sysname{} architecture.}
\label{fig:overview}
\vspace{-2mm}
\end{figure}

\sysname{} is a multi-model training system for large-scale models, co-designing inter-job dynamic scheduling and intra-job adaptive parallelism (AP) to achieve high cluster efficiency, scalability, and production practicality.
The key idea is to employ the two-fold approach (\S\ref{sec:challenges}) to acquire the AP performance of jobs beforehand, dynamically schedule them in an AP-aware manner, and execute them with efficient AP.

To hierarchically unify estimation and profiling, \sysname{} shards the joint optimization space into subspaces (``grids'') based on the observation in \S\ref{sec:challenges}.
A grid of a job includes all scheduling-parallelism plans with identical allocated resource and pipeline degree.
Within a grid, a near-optimal parallelism plan (``proxy plan'') is analytically estimated; among grids, proxy plans are profiled and used for scheduling.

Figure~\ref{fig:overview} depicts the \sysname{} architecture with four main components: planner, profiler, scheduler, and AP runtime to address the challenges (\S\ref{sec:challenges}).
Initially, users submit LMs to the job queue.
Given a queuing job, \textit{Parallelism Planner} (\textit{C\#1}) produces proxy plans via roofline-based plan generation and Pareto frontier deduction, in load-aware and execution-free manner (\S\ref{sec:estimation}).
For each proxy plan, \textit{Disaggregated Profiler} (\textit{C\#2}) uses model graph for operator disaggregation and profiles non-redundant operators on fragmented resources (usually a single GPU) (\S\ref{sec:profiling}).
Given profiled jobs, \textit{Dynamic Scheduler} (\textit{C\#3}) schedules in elasticity and heterogeneity dimensions via job launching and scaling mechanisms and a generalized event-driven policy (\S\ref{sec:sched}).
At runtime, jobs are executed via \textit{AP Runtime} with pruned search space (\S\ref{sec:runtime}).

\subsection{Sharding Space into Grids}\label{sec:grid}

\paragraph{Joint Optimization Space.}
In scheduling-parallelism co-design, the search space is formulated by the outer product of scheduling space $\mathcal{S}$ and adaptive parallelism space $\mathcal{P}$.
$\mathcal{S}$ is defined as $\{(J_i, n, m)\}$, where the $(J_i, n, m)$ scheduling plan indicates allocating job $J_i$ with $n$ GPUs of type $m$.
The parallelism space $\mathcal{P}$ is given by $\{({P}^{(s)}_\text{inter}, {P}^{(s)}_\text{intra}) \}$~\cite{alpa}, where ${P}^{(s)}_\text{inter}$ groups operators into $s$ stages and assigns GPUs for each, and ${P}^{(s)}_\text{intra}$ contains intra-stage parallelism for stages.
Consequently, the joint space is formulated as:
\begin{equation}
\mathcal{J} = \mathcal{S} \times \mathcal{P} = \{(J_i, n, m, {P}^{(s)}_\text{inter}, {P}^{(s)}_\text{intra}) | s \in [1,n] \}.
\label{eq:opt_space}
\end{equation}
Assume $K$ jobs, $O$ operators, $N$ maximum allocable GPUs and $M$ types, the overall complexity of $\mathcal{J}$ is $O(KNM \sum_s \binom{O}{s} \binom{N}{s} 2^s)$. 
Such a huge optimization space consists of innumerable ``scheduling-parallelism'' plans, calling for an efficient approach to identify performant ones.

\begin{figure}
\centering
\includegraphics[width=.98\linewidth]{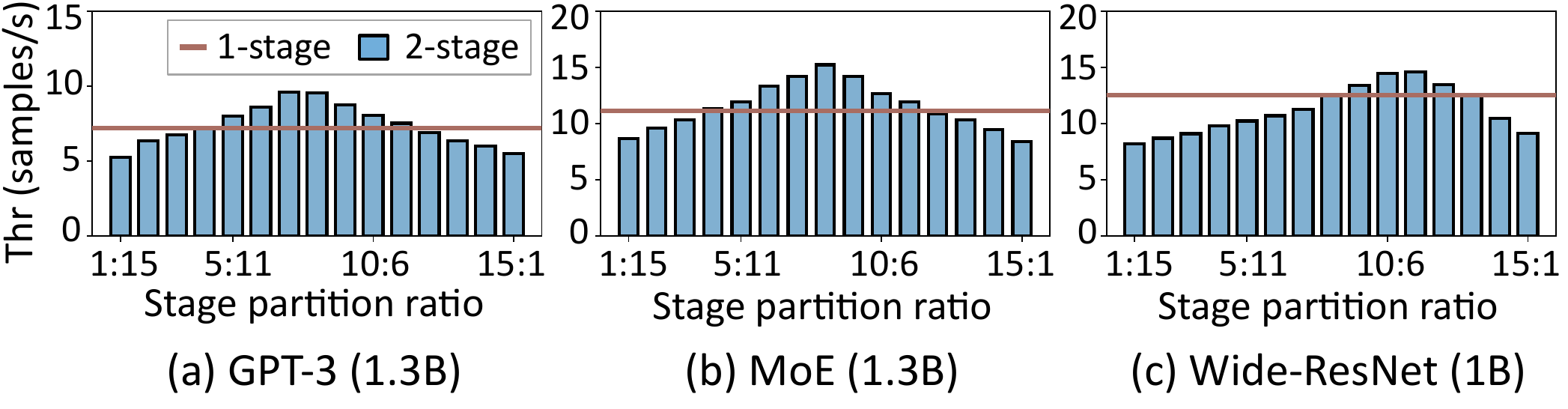}
\vspace{-3mm}
\caption{Throughput of various stage partition ratios compared to the single-stage case (with MP). \texttt{X:Y} represents the proportion of operators divided to the first and second stages. The later layers in \texttt{Wide-ResNet} are typically larger~\cite{wideresnet}.}
\label{fig:partition_ratio}
\vspace{-2mm}
\end{figure}

\begin{figure}
\centering
\includegraphics[width=.85\linewidth]{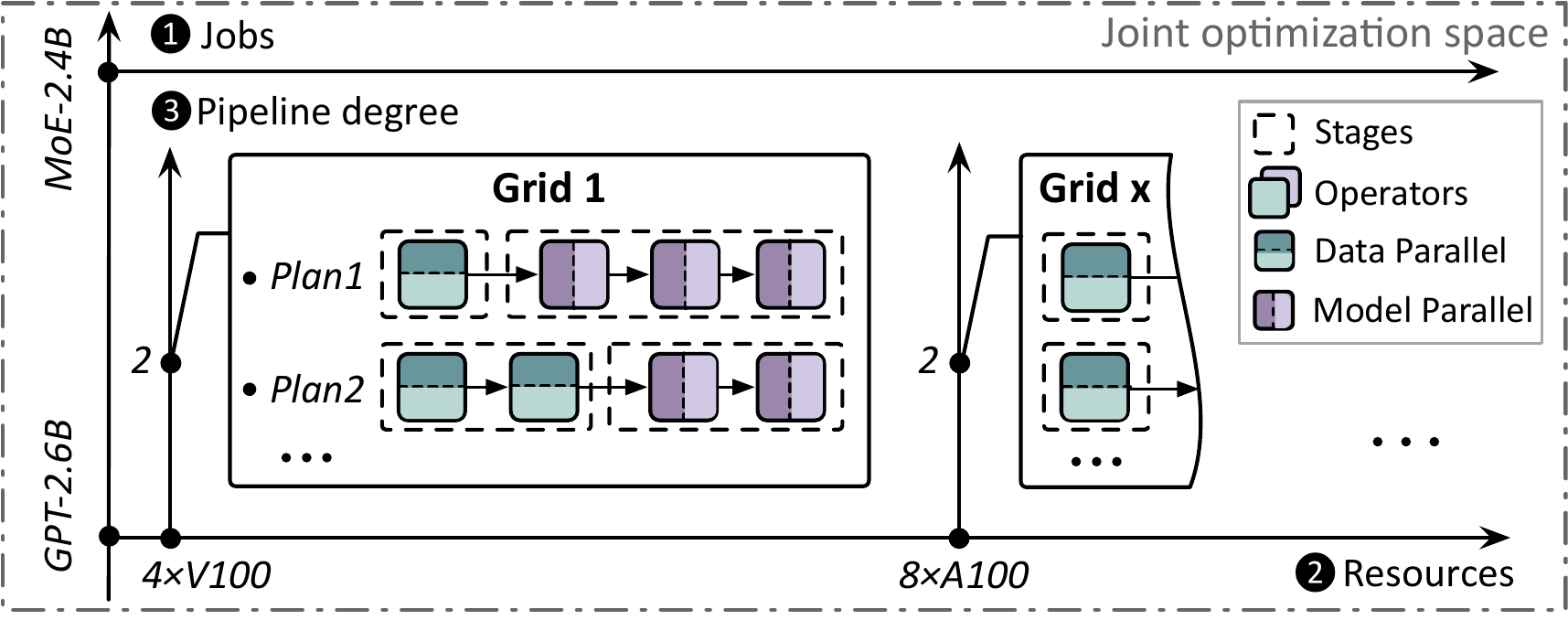}
\vspace{-2mm}
\caption{Sharding joint optimization space into grids.}
\label{fig:grid}
\vspace{-4mm}
\end{figure}

\paragraph{Granularity of Grid.}
For efficient exploration, the joint space is sharded into grids, i.e., $\mathcal{J} = \mathcal{J}_\text{out} \times \mathcal{J}_\text{in}$. 
Profiling is applied in $\mathcal{J}_\text{out}$ (among grids), whereas estimation is used in $\mathcal{J}_\text{in}$ (in each grid).
For instance, full-space profiling defines $\mathcal{J}_\text{out}$ as $\{(J_i, n, m, {P}^{(s)}_\text{inter}, {P}^{(s)}_\text{intra})\}$ and $\mathcal{J}_\text{in}$ as $\emptyset$; in Sia~\cite{sia}, $\mathcal{J}_\text{out}$ is given by $\{(J_i, m)\}$ and $\mathcal{J}_\text{in} = \{(n, {P}^{(s)}_\text{inter}, {P}^{(s)}_\text{intra})\}$.

To reconcile cost and accuracy, we determine the grid granularity based on the following observation.
For a model with fixed resources, when adaptively parallelizing it \textit{with a fixed pipeline degree}, a plan with balanced inter-stage loads consistently outperforms imbalanced ones on end-to-end performance (Figure~\ref{fig:partition_ratio}).
This is because each pipeline is bottlenecked by its slowest stage~\cite{computer_arch,pipedream}.
An intuitive counter example to show the essentiality of fixing the pipeline degree is that, a model with a single stage always exhibits perfect inter-stage balance, yet may underperform multi-stage cases (e.g., up to $1.34\times$ for GPT-3), as shown in Figure~\ref{fig:partition_ratio}(a)-(c).
Notably, training optimizations such as kernel fusion~\cite{tvm} and activation checkpointing~\cite{act-ckpt} have limited impact on this observation, as they uniformly reduce kernel launch overhead or memory footprint across pipeline stages.

This observation serves as the criterion of assessing parallelism plans without precise latency measurements.
Therefore, the grid of a job is defined as ``optimization subspace with determined resource and pipeline degree'' (Figure~\ref{fig:grid}). 
Concretely, $\mathcal{P}$ is dissected into $\{(s) \times ({O}_s, {D}_s, {P}^{(s)}_\text{intra})\}$, where ${O}_s$ and ${D}_s$ are stage partition and GPU assignment for $s$ stages.
$\mathcal{J}_\text{out}$ becomes $\{(J_i, n, m, s)\}$ and $\mathcal{J}_\text{in}$ is $\{({O}_s, {D}_s, {P}^{(s)}_\text{intra})\}$.
This sharding reduces the profiling complexity to $O(KN^2M)$, while ensuring the optimality of AP-aware estimation.

\subsection{Load-Aware Parallelism Planning}\label{sec:estimation}

Instead of relying on extensive operator profiling before parallelization~\cite{alpa,pipedream}, \sysname{} builds an execution-free parallelism planner that assesses plans in each grid by inter-stage load balancing and intra-stage cost minimization.

\begin{figure}
\centering
\includegraphics[width=0.95\linewidth]{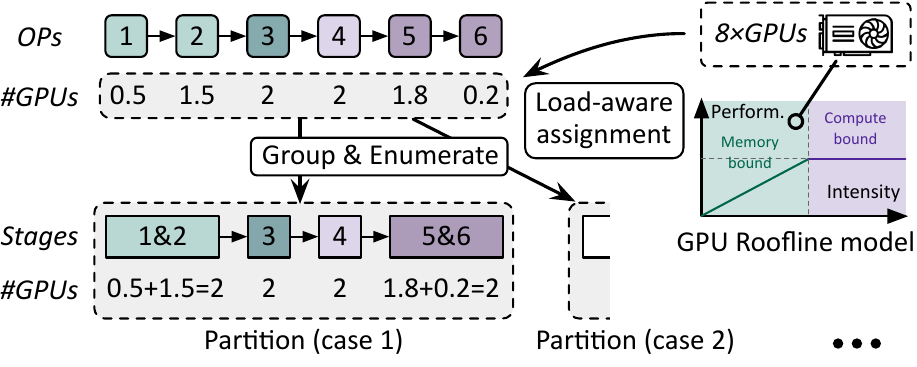}
\vspace{-1mm}
\caption{Example of stage partition and GPU assignment.}
\label{fig:plan_gen}
\vspace{-3mm}
\end{figure}

\paragraph{Roofline-Based Plan Generation.}
For a model with $O$ clustered operators\footnote{In practice, operators (e.g., matmul) are pre-clustered to control problem size (e.g., $O=16$~\cite{alpa}). We use the term ``operator'' to represent this.}, \sysname{} calculates operator FLOPs and memory access from static information (e.g., shapes).
Since operator performance is collectively affected by these two factors~\cite{fastermoe,habitat}, we define its \textit{load} as:
\begin{equation}
L_i = \frac{{FLOPs}_i}{R(I_i)} = \frac{{FLOPs}_i}{R({FLOPs}_i/{Mem}_i)},
\label{eq:load}
\end{equation}
where $I_i$ is the arithmetic intensity of operator $i$, determined by input and parameter settings (e.g., batch size). $R(\cdot)$ is the roofline model function, which models the attainable performance on specific devices~\cite{roofline} and depends only on hardware specifications (e.g., SM count, memory bandwidth).

\sysname{} then generates candidate stage partitions and GPU assignments.
Since intra-stage parallelism evenly divides a stage across GPUs~\cite{alpa,megatron}, the load of a parallelized operator is $L_i/d_i$, where $d_i$ is the GPU count.
Thus, with the criterion of load balancing, \sysname{} ensures \textit{each stage's GPUs are proportional to their loads}.
As shown in Figure~\ref{fig:plan_gen}, \sysname{} calculates $d_i$ for each operator by: $d_i = (L_i / \sum_{i'=1}^O L_{i'}) N_\text{gpu}$, where $N_\text{gpu}$ is the total number of GPUs.
After that, \sysname{} enumerates $\binom{O-1}{s-1}$ partitions by grouping operators into $s$ stages, and computes GPU assignments by: ${D}^{(j)}_s = \sum_{i} d_i, \ \forall OP_i \in Stage_j, \forall j \in [1, s]$.
In practice, ${D}^{(j)}_s$ is typically limited to a power-of-2~\cite{sia}, thus \sysname{} normalizes it by minimizing \textit{computation bias} metric based on Euclidean distance~\cite{eu_dist}:
\begin{equation}
b_\text{comp} = (\mathnormal{\sum}_{j=1}^s |{D}^{(j)}_{s,norm} - {D}^{(j)}_s|^2)^{\frac{1}{2}},
\label{eq:compute_dev}
\end{equation}
where ${D}^{(j)}_{s,norm}$ is the normalized GPU count for stage $j$. This metric quantifies the gap between the actual and theoretically optimal assignments. A larger value indicates a more imbalanced pipeline with poorer end-to-end performance.

To generate parallelism plans, \sysname{} further determines intra-stage parallelism per stage by minimizing communication cost within memory limits.
The observation is that, when evenly parallelizing stages, different intra-stage parallelism incurs varying communication and memory costs.
While sharding dimensions (e.g., batch dimension for DP) affect tensor layouts, communication dominates the selection with much larger latency impact (e.g., $6.27$ms v.s. $0.03$ms for a GEMM with $TP=4$/$DP=4$), as compute (SM) utilization is saturated under large-scale training workloads~\cite{alpa,metis}.


\paragraph{Deducing Pareto Frontier and Proxy Plan.}
Without execution, it is infeasible to define a unified metric (e.g., end-to-end latency~\cite{alpa,dynapipe}) that considers both computation and communication, hindering direct comparison between parallelism plans.
Therefore, \sysname{} evaluates computation and communication separately.
The computation performance is assessed by $b_{comp}$, which quantifies the impact of stage partitioning and GPU assignment.
For communication, we introduce a \textit{communication load} metric based on pipeline patterns (\S\ref{sec:profiling}) to evaluate the end-to-end cost:
\begin{equation}
l_\text{comm} = (B-1) \cdot \max_{1 \leq j \leq s} \{\sum_{i} c_i\} + \sum_{i'=1}^{O'} c_{i'}, \ \ \forall OP'_i \in Stage_j,
\label{eq:comm_bn}
\end{equation}
where $B$ is the number of microbatches, $OP'_i$ denotes communication operator $i$ and $c_i$ is the cost.
There are totally $O'$ communication operators. 
This metric models sequentially executed intra-stage communication (e.g., all-reduce), while excluding P2P communication as it is much smaller and usually overlapped with stage computation.

With the two metrics, \sysname{} frames plan comparison as a multi-objective optimization problem~\cite{pareto,multi_obj}: a parallelism plan is considered superior if it achieves both lower $b_\text{comp}$ and $l_\text{comm}$ compared to another plan. 
Thus, \sysname{} minimizes two metrics simultaneously, deducing a Pareto frontier by traversing parallelism plans and collecting \textit{non-dominated} ones (no other plan outperforms them on both metrics).
If the collected plans exceeds a threshold, the Pareto frontier is reduced by dropping the plan with higher communication load from the pair with the most similar stage partition.

Given the Pareto frontier, \sysname{} identifies the proxy plan from non-dominated plans.
Since computation typically dominates end-to-end performance~\cite{alpa,madmax}, \sysname{} filters plans with the minimum computation bias and then selects the plan with the lowest communication load, which achieves $93.4\%$ of the optimal plan performance on average (\S\ref{sec:plan_exp}).

\subsection{Disaggregated Profiling}\label{sec:profiling}

Though the grid sharding of \sysname{} significantly reduces profiling complexity (\S\ref{sec:grid}), considerable resources are still required to profile proxy plans, especially for LMs.
This incurs job queuing and other issues as analyzed in \S\ref{sec:strawman}.
To address this, we find that the operator execution after compilation remains sequential and isolated despite data dependencies~\cite{alpa,nnscaler}, thereby enabling operator-level profiling.
Existing solutions include zero-GPU blackbox prediction~\cite{neusight,habitat} with extensive pre-fitting and unstable errors on dynamic resources, and multi-GPU measurement~\cite{centimani,PCS} that requires full GPU access to benchmark operator latencies.

In contrast, \sysname{} designs a profiler for both precise and efficient profiling on fragmented GPUs (e.g., a single GPU), while generalizing to various runtime optimizations~\cite{tvm}.
In this subsection, we detail why and how \sysname{} disaggregates computation and communication operators, then present the profiling approach using a single device.


\paragraph{Operator Disaggregation.}
We observe \textit{compute redundancy} when profiling parallelized LMs, including:
 (\romannumeral1) Workers of the same stage execute identical operators~\cite{alpa,nnscaler}, while the stage latency equals that of a single worker;
 (\romannumeral2) Many LMs contain a large proportion of repeated operators with identical latency~\cite{gpt-3,llama2};
 (\romannumeral3) Device stalls (e.g., pipeline bubbles~\cite{gpipe}) largely exist during model execution.

To eliminate unnecessary computations and device stalls, \sysname{}  disaggregates the operators of a model.
Specifically, \sysname{} pre-compiles the functions of each parallelized stage and generates intermediate representation (IR) objects. 
Then, \sysname{} extracts computation graphs from IR objects and retrieves operator configurations (e.g., shapes, data types) from them.
The disaggregated computation and communication operators are taken into separated consideration:



\textbf{\textit{1)} Computation operators.}
The latency of a computation operator is affected by multiple factors.
First, different kernel implementations with irregular latencies are launched for diverse shapes, data types, and GPU architectures~\cite{habitat,sarathi_serve}. 
Second, various runtime optimizations (e.g., operator fusion) are adopted to optimize execution~\cite{tvm,rammer,xla_analysis}.
Thus, online profiling is essential to obtain the precise latency of computation operators, and can be done on a single GPU.

\textbf{\textit{2)} Communication operators.}
The latency of a communication operator depends on primitive types, network topologies (fixed after hardware setup), and transfer volume.
Advanced optimizations such as primitive substitution and data compression also statically rely on primitive types and GPU interconnection~\cite{centauri,alpa,flash_comm}.
Thus, under the same primitive and network topology, the latency of a communication operator is proportional to data transfer volume~\cite{torch_comm,fastermoe,madmax}.

\begin{figure}
\centering
\includegraphics[width=0.95\linewidth]{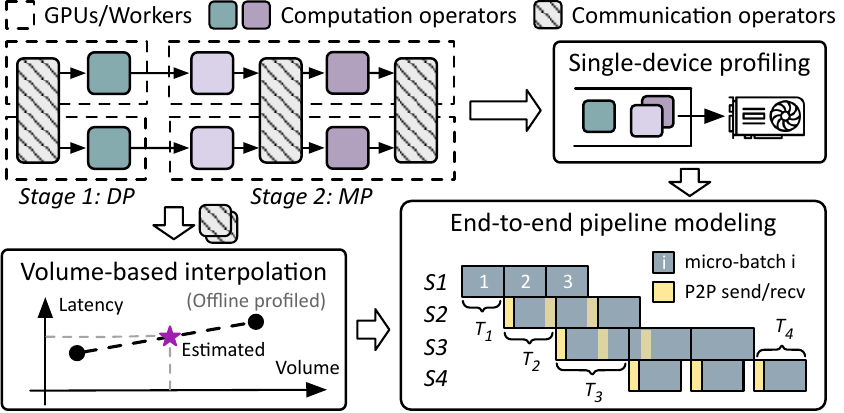}
\vspace{-1mm}
\caption{Workflow of single-device disaggregated profiling.}
\label{fig:profile}
\vspace{-2mm}
\end{figure}

\paragraph{Single-Device Profiling.}
\sysname{} dispatches computation and communication operators to dedicated modules (Figure~\ref{fig:profile}).
For computation, \sysname{} reconfigures operator shapes based on the intra-stage parallelism (e.g., dividing batch dimension by 2 for 2-way DP).
For each stage, \sysname{} constructs a single-device executable with reconfigured operators, corresponded to a worker of that stage. 
Lastly, \sysname{} profiles executables with the same runtime optimizations as in direct execution (thus achieving kernel-level equivalence), and captures the latencies of launched kernels~\cite{cupti}.
Operators with identical configurations are not profiled repeatedly.

For communication, \sysname{} offline samples representative data volumes and profiles candidate primitives across pre-accessible hardware (e.g., $2\times4$ A100 nodes with Infiniband).
At online, \sysname{} estimates operator latency by interpolating based on the transfer volume under the same primitive and hardware.
The stage latency is calculated by summing the latencies of all operators due to sequential execution.

Finally, \sysname{} models the execution schedule of specific pipeline (e.g., 1F1B~\cite{pipedream}) and computes the end-to-end latency $T_\text{e2e}$.
As shown in the bottom right of Figure~\ref{fig:profile}, $T_\text{e2e}$ consists of: 
 (\romannumeral1) Latencies of the first microbatch for all stages ($T_1+T_2+T_3+T_4$);
 (\romannumeral2) Latencies of the remaining $B-1$ microbatches for the slowest stage ($(B-1) \times (T_3 - T_\text{p2p})$), where computation-communication overlap is modeled.

\subsection{Dynamic Scheduling}\label{sec:sched}


To query the AP performance of a job on specific resources, \sysname{} traverses relevant grids for the best-performing one, using its profiled data as inputs to the scheduler.




\paragraph{Mechanisms.}
\sysname{} provides two fundamental ``operations'' during job launching and training phases.

\textbf{\textit{1)} Priority-Based Multi-Queue Launching.}
In \sysname{}, jobs are assigned priority $\lambda \in [1, P]$ based on user or vendor specifications (aligned with production practice~\cite{acme,antman}).
\sysname{} maintains $P$ queues to schedule job launching order (one for each $\lambda$ value).
A smaller $\lambda$ indicates higher launching priority, while jobs within the same queue are launched in enqueue order.
A launched job adheres to the ``run-to-completion'' rule, i.e., cannot be preempted or scaled to a non-executable state due to high migration overhead of large models.
To mitigate starvation caused by head-of-line blocking~\cite{jeon2019analysis,antman}, when a job blocks $\lambda$-queue due to substantial resource demands, a subsequent job in $\lambda$-queue is allowed to preempt resources (when feasible) for earlier launch — a privilege not for jobs in $(\lambda+1)$-queue.
For fairness, a job priority $\lambda$ is promoted to $(\lambda - 1)$ (if $\lambda$ > 1) after prolonged queuing.
The selection of $P$ value is analyzed in \S\ref{sec:search_depth}.


\textbf{\textit{2)} Two-Dimensional Scaling.} 
To improve cluster efficiency and reduce fragmentation, \sysname{} lively \textit{scales} resources for both newly launched and in-flight jobs, by adjusting the allocated number (``elasticity'') and type (``heterogeneity'') of GPUs.
To ensure job locality (or rack affinity), \sysname{} follows the buddy allocation rule and attempts periodic job migration for defragmentation~\cite{elasticflow,hived}.
User-specified hyperparameters (e.g., global batch size) remain fixed to ensure that the scaling does not compromise model quality (\S\ref{sec:search_depth}).

Notably, \sysname{} employs dynamic GPU type switching to reduce job queuing and improve resource utilization, while ensuring intra-job homogeneity to guarantee training efficiency and ease of management~\cite{sia,gavel,easyscale}.
This is because production clusters typically house homogeneous GPUs in the same region with neighboring nodes. Allocating heterogeneous GPUs to a single job results in cross-region communication with much limited bandwidth~\cite{heteroscale,helios,diloco}.
Moreover, due to the risks of compute stragglers and resource fragmentation, intra-job heterogeneity is more considered in cloud-based training, where large-block GPU allocation in a single region is often infeasible~\cite{crosspipe,cross_region}.

\begin{algorithm}[t]
\begin{algorithmic}[1]
\small

\State \textbf{Global}: In-flight jobs $J_{L}$, priority queue $Q$, cluster resources $R$, search depth $D$

\Function{Schedule}{$sEvents, cEvents$}  \Comment{\textcolor{gray}{Entrypoint}}
    \State $R \leftarrow$ \Call{Free}{$cEvents.resources$} \Comment{\textcolor{gray}{Free resources}}
    \State $J_L \leftarrow J_L - cEvents.jobs$ 
    
    \State $Q.\text{enqueue}(sEvents)$; $plans \leftarrow \emptyset$; $\lambda' \leftarrow P + 1$
    \For{$E \in Q.\text{sort()}$}
        \State $job \leftarrow E.job$; $plans[job] \leftarrow$ \Call{LEventHandler}{$E$}
        \If{$ plans[job] \textbf{ is } \emptyset $\textbf{ and} $job.\lambda \leq \lambda'$} $\lambda' \leftarrow job.\lambda$
        \EndIf
        \If{$ plans[job] \textbf{ is } \emptyset $\textbf{ and} $job.\lambda > \lambda'$} \textbf{break}
        \EndIf
    \EndFor
    \State $plans \leftarrow plans +$ \Call{InFlightHandler}{$ $}
    \State \Call{Finalize}{$plans, J_L, R$}  \Comment{\textcolor{gray}{Allocate resources}}
\EndFunction

\Function{LEventHandler}{$Event$}  \Comment{\textcolor{gray}{Launch a job}}
    \While{\textbf{not} $Event.alloc$ \textbf{and} $plan.len < D$}
        \State $plan \leftarrow$ \Call{GetOptimalScaleDown}{$J_L + Event.job, R$}
    \EndWhile
    \If{$Event.alloc$} $J_L \leftarrow J_L + Event.job$; $Q.\text{dequeue}(Event)$
    \EndIf
    \State \Return{$plan$} \textbf{if} $Event.alloc$ \textbf{else} $\emptyset$
\EndFunction

\Function{InFlightHandler}{$ $}  \Comment{\textcolor{gray}{Scale up in-flight jobs}}
    
    \While{$job^*$ \textbf{is not} $\emptyset$ \textbf{and} $plan.len < D$}
        \State $job^*, OP^* \leftarrow$ \Call{GetOptimalScaleUp}{$J_L, R$}
        \If{$job^*$ \textbf{is not} $\emptyset$}
            $plan \leftarrow plan + <job^*, OP^*>$
        \EndIf
    \EndWhile
\EndFunction

\caption{Generalized Event-Driven Job Scheduling}
\label{alg:schedule}
\end{algorithmic}
\end{algorithm}

\paragraph{Generalized Event-Driven Scheduling.}
\sysname{} employs a generalized event-driven policy with support of diverse scheduling objectives.
For instance, the throughput maximization objective is formulated as:
\begin{equation}
\max_{A} \sum_{i=1}^K (\sum_{m=1}^{M}\sum_{n=1}^{N} A_i[m, n]  \times Thr^{m,n}_{i}), 
\label{eq:thr_max}
\end{equation}
where $A_{i}[m, n]=1$ denotes allocating job $J_i$ with $n$ GPUs of type $m$.
$Thr^{m,n}_{i}$ is the AP throughput.
Other intuitive constraints are not shown due to space limits.
For deadline awareness, \sysname{} adds an additional constraint:
\begin{equation}
\forall i, \sum_{m=1}^{M}\sum_{n=1}^{N} \frac{A_i[m, n]  \times B_iN^{iter}_i}{Thr^{m,n}_i} \leq DDL_{i},
\label{eq:ddl}
\end{equation}
where $B_i$, $N^{iter}_i$, and $DDL_i$ denote the global batch size, remained iteration count, and deadline of job $J_i$.

Algorithm~\ref{alg:schedule} outlines the scheduling workflow of \sysname{} with three job event types: submission (\texttt{sEvent}), launch (\texttt{lEvent}), and completion (\texttt{cEvent}).
In each scheduling step, \sysname{} invokes \texttt{Schedule} to handle pending \texttt{sEvent}s and \texttt{cEvent}s (line 2-11). 
For \texttt{lEvent}, \sysname{} calls \texttt{LEventHandler} to iteratively locate the optimal job to scale down based on the scheduling objective (line 12-18).
If idle resources exist, \sysname{} then calls \texttt{InFlightHandler} to iteratively locate the optimal in-flight job to scale up (line 19-24).

To efficiently locate the optimal job for scaling, \sysname{} heuristically identifies the job that yields the greatest benefit after scaling. 
This is backed by the observation in \S\ref{sec:contradiction} that for different models and hardware, scaling operations lead to varied performance variations with diminishing returns.
Therefore, \sysname{} scales down jobs with excessive resources but limited performance when resources are insufficient, and scales up promising jobs if idle resources exist.
Prior schedulers have demonstrated the near-optimality of such heuristics~\cite{elasticflow,easyscale}.
To reduce scheduling overhead, a parameter named \textit{search depth} limits the maximum iterations in the iterative scaling process (evaluated in \S\ref{sec:search_depth}).

\paragraph{Extensibility.}
Beyond the throughput maximization and deadline awareness, \sysname{} is compatible to other objectives that allocate resources based on job performance.
For example, the Finish-Time Fairness~\cite{mahajan2020themis} can be formulated as:
\begin{equation}
\min \max_{1 \leq i \leq K}(\sum_{m=1}^{M}\sum_{n=1}^{N} \frac{N^{iter}_i T^{m,n}_i}{A_i[m,n] \times Q^{m,n}}),
\label{eq:fairness}
\end{equation}
where $T^{m,n}_i$ is the iteration time and $Q^{m,n}$ measures the allocated resources.
In addition, the scheduling policy can be modified to seamlessly integrate with other algorithms, such as solving Equation~\ref{eq:thr_max} by ILP optimization. 
\subsection{Runtime Pruning}\label{sec:runtime}

At runtime, \sysname{} executes scheduled jobs with adaptive parallelism (AP), pruning the search space via three predefined rules for fast deployment.
First, for each job with allocated resources, \sysname{} selects the best-performing grid ($G_b$) and directly applies its pipeline degree.
Second, \sysname{} identifies the most imbalanced stage partition in the Pareto-optimal plans of $G_b$, pruning plans with greater imbalance.
Third, for each stage, if its operator composition matches that of a stage in certain Pareto-optimal plans, its assigned GPU count and intra-stage parallelism are directly determined.
These rules are compatible with various AP frameworks (e.g., Alpa~\cite{alpa}, Aceso~\cite{aceso}, nnScaler~\cite{nnscaler}).
Alternatively, \sysname{} supports directly using the proxy plan for zero-overhead execution.
Experiments in \S\ref{sec:plan_exp} demonstrate $5.48\times$ search cost reduction ($96.2\%$ of the optimal performance) on average.


\section{Implementation}\label{sec:impl}

We implement \sysname{} with 13K LoC in Python: 2,800 lines for the parallelism planner, 5,600 lines for the disaggregated profiler, and 5,200 lines for the scheduler.
\sysname{} provides a simulator to conduct large-scale scheduling experiments, ensuring high fidelity by sharing scheduling codes and logics with the real-testbed scheduler.
We build the training backend based on Alpa~\cite{alpa} in JAX framework~\cite{jax}, with minor modification of 500 LoC on its \texttt{stage\_construction} and \texttt{stage\_profiling} modules to support parallelism pruning at runtime.
Besides such modifications, \sysname{} modules remain independent of the underlying backend.

For parallelism planning, we obtain operator information (e.g., FLOPs, memory access) via static model analysis based on XLA~\cite{xla} and its HLO IR.
We borrow the ILP solver of Alpa to minimize communication costs in intra-stage parallelization phase.
For disaggregated profiling, stage pre-compilation is implemented based on XLA. We implement a kernel profiler with 500 LoC in C++ based on NVIDIA CUPTI~\cite{cupti}, using it to capture kernel latencies during single-device profiling.
We offline profile representative communication primitives (e.g., all-reduce) based on XLA, NCCL~\cite{NCCL} and Ray~\cite{moritz2018ray}.
For scheduling, we use gRPC~\cite{grpc} to communicate between the scheduler and distributed training processes.
\sysname{} schedules job launch and completion events every 5 minutes and inspects cluster status every 20 seconds.
\sysname{} implements job migration operations based on checkpoint-resume~\cite{gandiva}, while remaining extensible to advanced state management techniques~\cite{bytecheckpoint,tenplex}, such as delegating model and dataset states to Tenplex~\cite{tenplex} for faster model migration and reinitialization.



\section{Evaluation}\label{sec:eval}

\begin{table}[t]
\small
\centering
\caption{Specifications of heterogeneous GPU clusters in evaluation. GPUs with $^{\dagger}$ are equipped with NVLink~\cite{nvlink}.}
\vspace{-2mm}
\begin{tabular}{c@{\hspace{0.2cm}}c@{\hspace{-0.05cm}}c@{\hspace{-0.00cm}}c@{\hspace{-0.00cm}}c}
\toprule
GPU & Arch. & Mem(GB) & Interconnect & \#GPU/Node \\
\midrule
H100$^{\dagger}$ & Hopper~\cite{Hopper} & 80 & NV. ConnectX-6~\cite{cx6} & 8 \\
L20 & Ada~\cite{Ada} & 48 & NV. ConnectX-6 & 16\\
A100$^{\dagger}$ & Ampere~\cite{Ampere} & 40 & NV. ConnectX-5~\cite{cx5} & 4\\
A40 & Ampere & 48 & NV. ConnectX-5 & 2\\
A10 & Ampere & 24 & NV. ConnectX-6 & 2\\
V100$^{\dagger}$ & Volta~\cite{Volta} & 32 & NV. ConnectX-5 & 16\\
\bottomrule
\end{tabular}
\label{tab:sim_cluster}
\vspace{-1mm}
\end{table}

\subsection{Experimental Setup}\label{sec:setup}

\paragraph{Physical Testbeds.}
We use two heterogeneous clusters:
(\romannumeral1) \textit{Cluster-A}: A cluster of 32 nodes and 64 GPUs. 16 nodes have Intel Xeon Gold 5318Y CPUs and 2 NVIDIA A40 GPUs; other 16 nodes have the same CPUs but with 2 A10 GPUs.
(\romannumeral2) \textit{Cluster-B}: A cutting-edge cluster with 128 NVIDIA H100 GPUs (16 nodes, each with 2 Intel Xeon Platinum 8457C CPUs and 900GB/s NVLink) and 256 L20 GPUs (16 nodes, each with the same CPUs and 64GB/s PCIe4.0 connection).



\paragraph{Simulated Cluster with Higher Heterogeneity.}
The simulated cluster has 1,280 GPUs of 4 types: A100 (80 nodes), A40 (160 nodes), A10 (160 nodes), and V100 (20 nodes), as detailed in Table~\ref{tab:sim_cluster}. 
For simulation, we extensively pre-measure AP performance across diverse models, hyperparameters, and hardware, replacing real model training with process sleeping and querying pre-measured data.
The measurement naturally includes the impact of network topologies (e.g., NVLink/Infiniband within/across A100 nodes).
We model the overhead of full-space/space-pruned AP and checkpoint-resume by pre-measuring across model and hardware scales.


\paragraph{Models and Traces.}
We conduct experiments using three LMs in Table~\ref{tab:models}.
For GPT-3 and MoE, we use the sequence length of 1024.
The number of micro-batches is set to $4\times$ the number of pipeline stages~\cite{gpipe}.
We use three production traces to evaluate the scheduling performance, including a two-week Philly trace with over 13,000 jobs~\cite{jeon2019analysis}, a Helios Venus trace~\cite{helios}, and a PAI trace~\cite{pai}.
For each job record (ID, submission time, duration), we randomly generate GPU count, type, model configurations, and iteration count to adapt the traces to heterogeneous scenarios.

\begin{table}[t]
\small
\centering
\caption{Model configurations used in the experiments.}
\vspace{-2mm}
\begin{tabular}{ccc}
\toprule
Model & Global Batch Size  & \#Params (billion)\\
\midrule
Wide-ResNet~\cite{wideresnet} & [256, 512, 1024] & [0.5, 1.0, 2.0, 4.0, 6.8]\\
GPT-3~\cite{gpt-3} & [128, 256, 512] & [0.76, 1.3, 2.6, 6.7]\\
GShard MoE~\cite{gshard} & [256, 512, 1024] & [0.69, 1.3, 2.4, 10, 27] \\
\bottomrule
\end{tabular}
\label{tab:models}
\vspace{-2mm}
\end{table}

\paragraph{Baselines.}
We compare \sysname{} against four baselines:
(\romannumeral1) \textbf{FCFS}~\cite{kube} rigidly schedules jobs with user-specified resources in arrival order.
(\romannumeral2) \textbf{Gavel}~\cite{gavel} performs heterogeneity-aware scheduling by formulating various objectives (e.g., throughput maximization) into ILP-based optimization problems and solving them.
(\romannumeral3) \textbf{ElasticFlow} (EF)~\cite{elasticflow} heuristically schedules jobs by scaling their GPU counts with deadline awareness and throughput maximization in a homogeneous cluster. 
(\romannumeral4) \textbf{Sia}~\cite{sia} schedules statically parallelized jobs, adjusting GPU counts, types, and hyperparameters by solving an ILP-based goodput maximization problem. 
Other RL-based schedulers~\cite{aware} are not used as their performance upper bounds would not exceed our ILP-based baselines.

\paragraph{Job Execution and Profiling.}
For baselines designed for SP-aware scheduling, we enable adaptive parallelism (AP) in job execution.
Once admitted, a job cannot be suspended by others.
Moreover, in physical and simulated testbeds, all dynamic schedulers require profiling results to schedule jobs.
We follow~\cite{elasticflow,lucid,gavel} to conduct all profiling ahead-of-time due to budget constraints, including full-GPU DP profiling for baselines, bootstrapped DP profiling for Sia, and grid-based techniques for \sysname{}. 
We prepend corresponded profiling overheads in scheduling experiments, which are directly measured in ahead-of-time profiling across models and hardware scales. 
The effectiveness of the \sysname{} planner and profiler is exclusively evaluated in \S\ref{sec:plan_exp} and \S\ref{sec:prof_exp}.

\subsection{Evaluation in Real-World Clusters}\label{sec:physical_exp}

\begin{figure}[t]
\centering
\includegraphics[width=\linewidth]{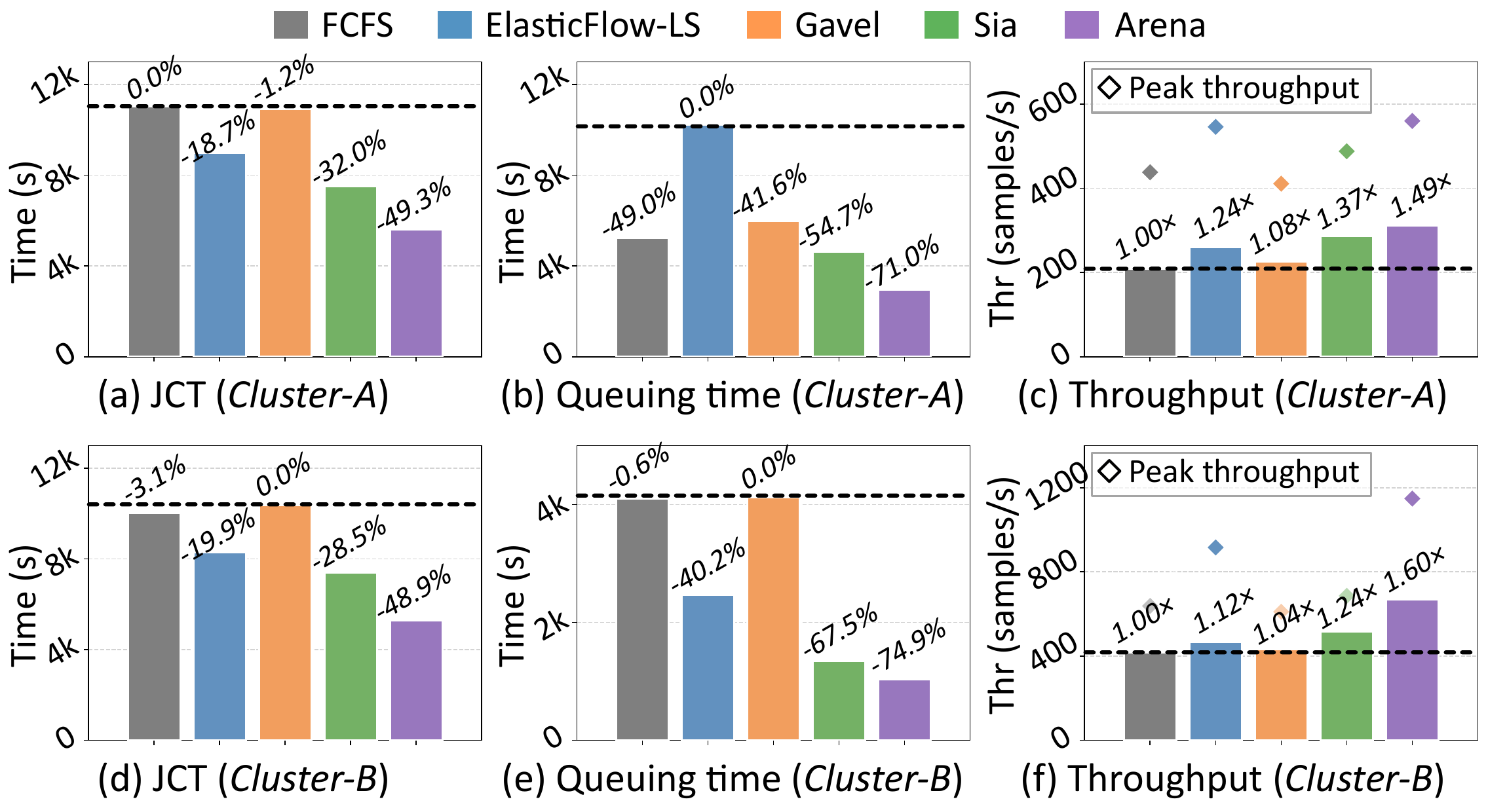}
\vspace{-6mm}
\caption{
Performance on two real-world testbeds.
\texttt{-LS} denotes ElasticFlow uses loosened job deadline~\cite{elasticflow}.
}
\label{fig:physical_exp}
\vspace{-2mm}
\end{figure}

We first evaluate \sysname{} on two heterogeneous testbeds (\S\ref{sec:setup}) using a 6-hour trace of 244 jobs from Philly trace~\cite{jeon2019analysis}. 
For \textit{Cluster-B} that is equipped with newer-generation GPUs, we scale up the workload by increasing model sizes and number of iterations (by $10\times$ on average) in the job trace.

In \textit{Cluster-A} (Figure~\ref{fig:physical_exp}(a)-(c)), \sysname{} improves average cluster throughput by up to $1.49\times$, reduces average JCT of all submitted jobs by $49.3\%$ and average queuing time by $71.0\%$.
The benefits stem from two main reasons.
First, \sysname{} identifies and profiles near-optimal parallelism plans for AP-aware job scheduling, achieving wiser resource allocation (e.g., mitigating overestimation issues in \S\ref{sec:contradiction}) and inter-job adjustment (e.g., allocating more GPUs to those with higher performance potential).
Second, \sysname{} reduces the time and hardware costs for job profiling and conditionally preempts blocked jobs, bringing better resource utilization for training and more execution opportunities.

In \textit{Cluster-B} with more advanced GPUs and heavier workloads (Figure~\ref{fig:physical_exp}(d)-(f)), \sysname{} also achieves significant improvements: up to $1.60\times$ higher cluster throughput, $47.3\%$ JCT reduction, and $74.9\%$ queuing time reduction.
Similar to the results in \textit{Cluster-A}, the awareness of AP performance in job scheduling acts as the primary reason for the benefits.
For baselines, though using more and better GPUs (thereby reducing JCT and queuing time), profiling larger models using DP with replicated parameters exacerbates the resource quota overestimation for AP execution.
Moreover, the enlarged cluster size also leads to larger scheduling space with increased resource misallocation risks.

\paragraph{Simulation Fidelity.}
We conduct simulations using the same configurations as real-world experiments in \textit{Cluster-A}. The average simulation error across \sysname{} and baselines is $3.16\%$ for throughput and  $7.22\%$ for JCT, which clearly confirms the fidelity of our forthcoming large-scale simulations.


\subsection{Larger-Scale Simulations}\label{sec:sim}



\begin{figure}[t]
\centering
\includegraphics[width=\linewidth]{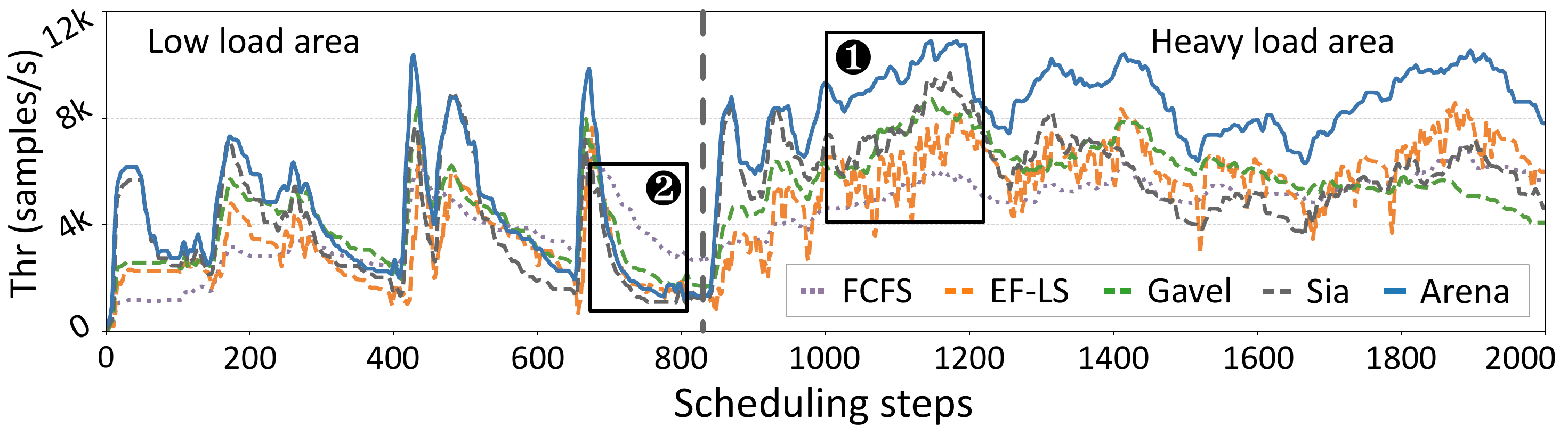}
\vspace{-6mm}
\caption{In-depth analysis of cluster throughput in 1,280-GPU simulated cluster. The scheduling interval is 5 minutes.
}
\label{fig:overall_thr}
\vspace{-1mm}
\end{figure}

Figure~\ref{fig:overall_thr} illustrates the cluster throughput of \sysname{} and the baselines with the one-week Philly trace.
In the first three days, the cluster load is low with several transient bursts; in last four days, the cluster experiences intensive heavy loads.
\sysname{} outperforms baselines when scaling up under burst loads because of better scheduling plans with more admitted jobs (\ding{182}). 
When loads decrease, \sysname{} scales down earlier as jobs are efficiently completed (\ding{183}).

\begin{figure}[t]
\centering
\includegraphics[width=\linewidth]{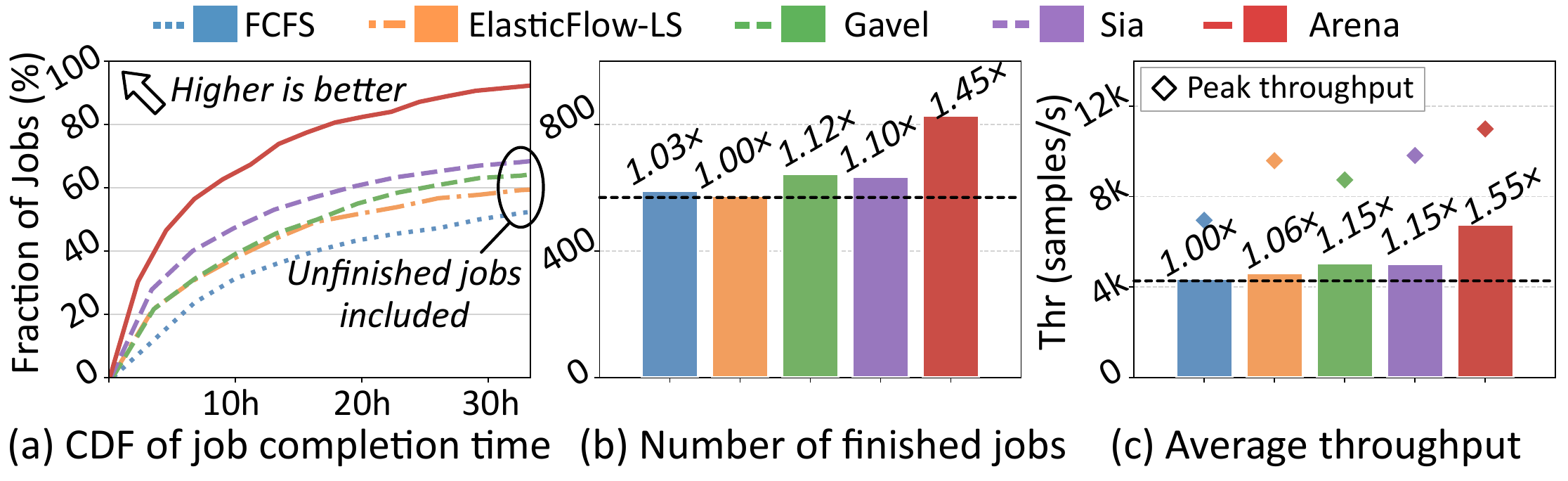}
\vspace{-6mm}
\caption{Performance analysis in the simulated cluster.}
\label{fig:sim_exp}
\vspace{-1mm}
\end{figure}

Figure~\ref{fig:sim_exp} further shows numerical comparisons of the above experiment.
As observed, \sysname{} reduces average JCT by $81.3\%$ (FCFS), $80.5\%$ (ElasticFlow-LS), $76.6\%$ (Gavel) and $75.2\%$ (Sia), completing up to $1.45 \times$ more jobs.
From the cluster perspective, \sysname{} outperforms baselines with up to $1.55 \times$ higher average throughput and $1.58 \times$ higher peak throughput.
Notably, the average job rescheduling count in \sysname{} is only $2.29$, as \sysname{} schedules based on precise AP performance and limits the search depth of job scaling (3 in simulation).
Gavel outperforms ElasticFlow on JCT and throughput, contrary to real-testbed results due to higher heterogeneity of the simulated testbed (\S\ref{sec:ablation}).
Sia outperforms other baselines under low loads, yet is throttled when cluster load bursts (Figure~\ref{fig:overall_thr}), as it schedules with DP and overestimates the resource demands of jobs.


\begin{figure}[t]
\centering
\includegraphics[width=.98\linewidth]{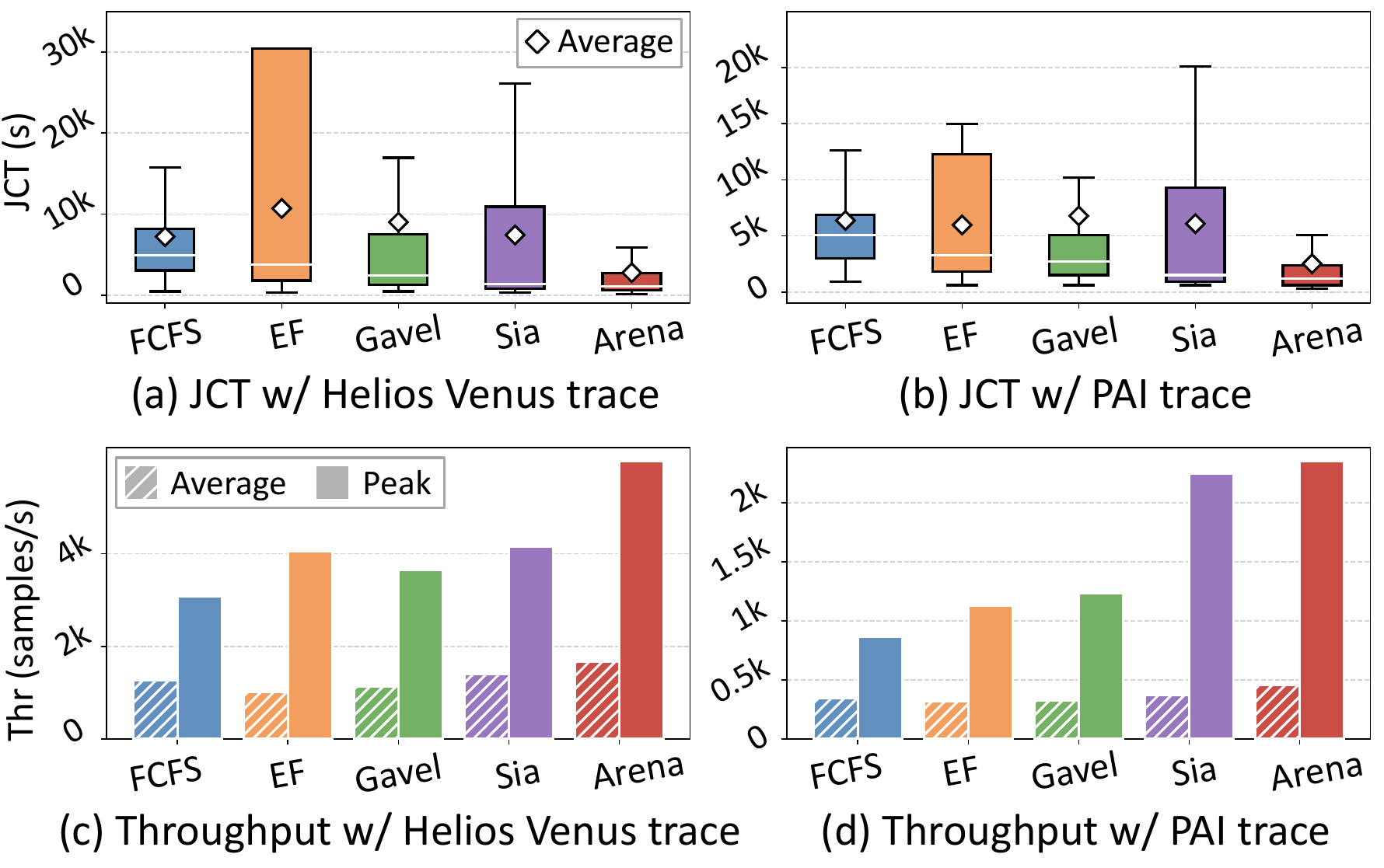}
\vspace{-2mm}
\caption{Performance on Helios (moderate loads) and PAI trace (light loads). Other settings are the same as before.}
\label{fig:other_traces}
\vspace{-2mm}
\end{figure}

\paragraph{Evaluation with Other Traces.}
We use a one-day trace with moderate load from Helios~\cite{helios} and another with low load from PAI~\cite{pai}.
Figure~\ref{fig:other_traces} shows that \sysname{} consistently outperforms baselines, achieving up to $74.2\%$ and $63.0\%$ JCT reduction, as well as $1.64 \times$ and $1.44 \times$ average throughput improvements on Helios and PAI traces.
These improvements are consistent with the results of Philly trace experiments.

\subsection{Effectiveness of Parallelism Planning}\label{sec:plan_exp}

\begin{figure}
\centering
\includegraphics[width=.98\linewidth]{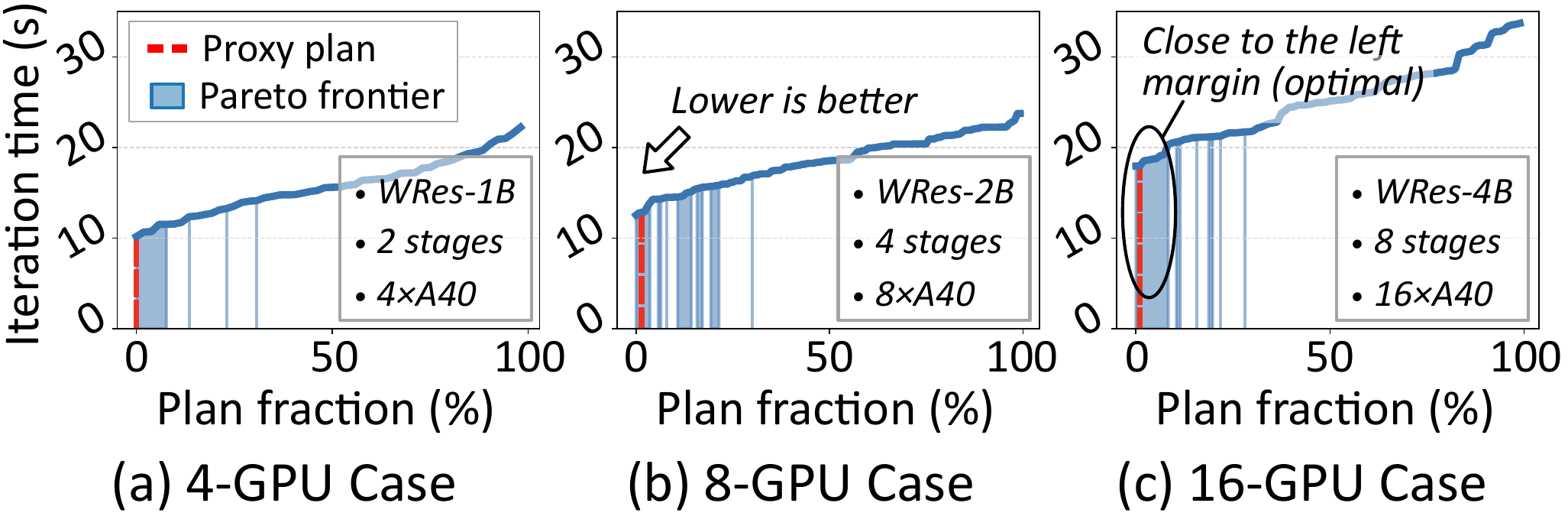}
\vspace{-2mm}
\caption{In-depth case study of Pareto frontier deduction.}
\label{fig:deduction}
\vspace{-1mm}
\end{figure}

\paragraph{Pareto Frontier and Proxy Plan Deduction.}
We evaluate the optimality of parallelism deduction by three case studies (Figure~\ref{fig:deduction}). 
Due to space limits, all experiments employ Wide-ResNet of varying sizes (most complex layer structure).
To build performance curves, we enumerate, profile, and sort all parallelism plans by iteration time in ascending order.

In each grid, the proxy plan achieves $86.2\%$, $85.6\%$, and $94.3\%$ of the optimal performance on 4, 8, and 16 GPUs.
Extending to more grids, the proxy plan matches the optimal plan in $86.5\%$ cases, with average performance gap of $9.82\%$ in mismatched cases.
Moreover, across various models and hardware, the best proxy plan (used for scheduling) achieves average $93.4\%$ performance of the AP searched optimal plan.

\begin{figure}
\centering
\includegraphics[width=.98\linewidth]{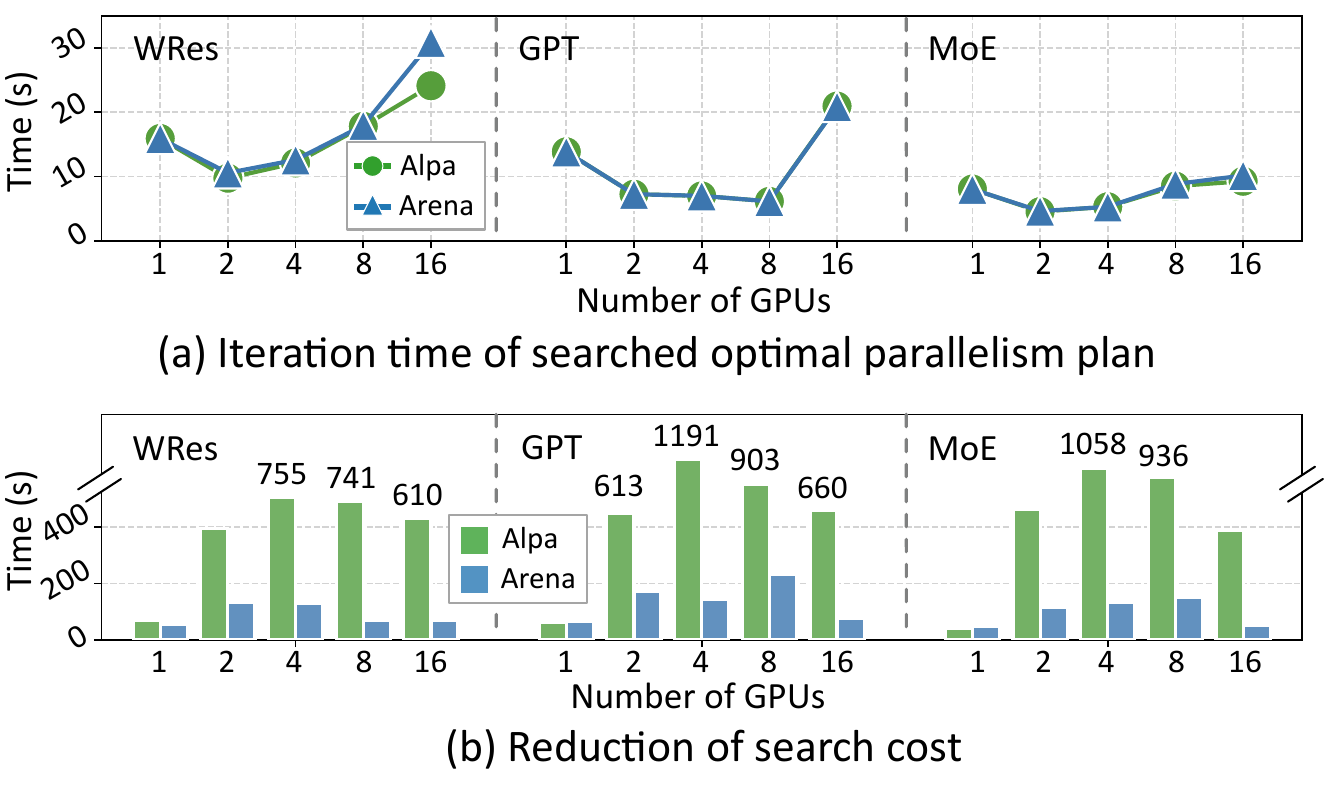}
\vspace{-2mm}
\caption{Performance of AP with pruning.}
\label{fig:pruning}
\vspace{-2mm}
\end{figure}

\paragraph{Parallelism Pruning.}
We further evaluate the effectiveness of parallelism pruning against Alpa~\cite{alpa}, as shown in Figure~\ref{fig:pruning}.
The optimal parallelism plan searched by \sysname{} achieves $96.2\%$ of Alpa performance on average, benefited from the optimality of Pareto frontier.
With pruning, \sysname{} reduces the AP search cost by $5.48\times$ on average and up to $10.88\times$, owing to the efficient pruning rules as in \S\ref{sec:runtime}.


\subsection{Efficiency of Disaggregated Profiling}\label{sec:prof_exp}

\begin{figure}[t]
\centering
\includegraphics[width=.98\linewidth]{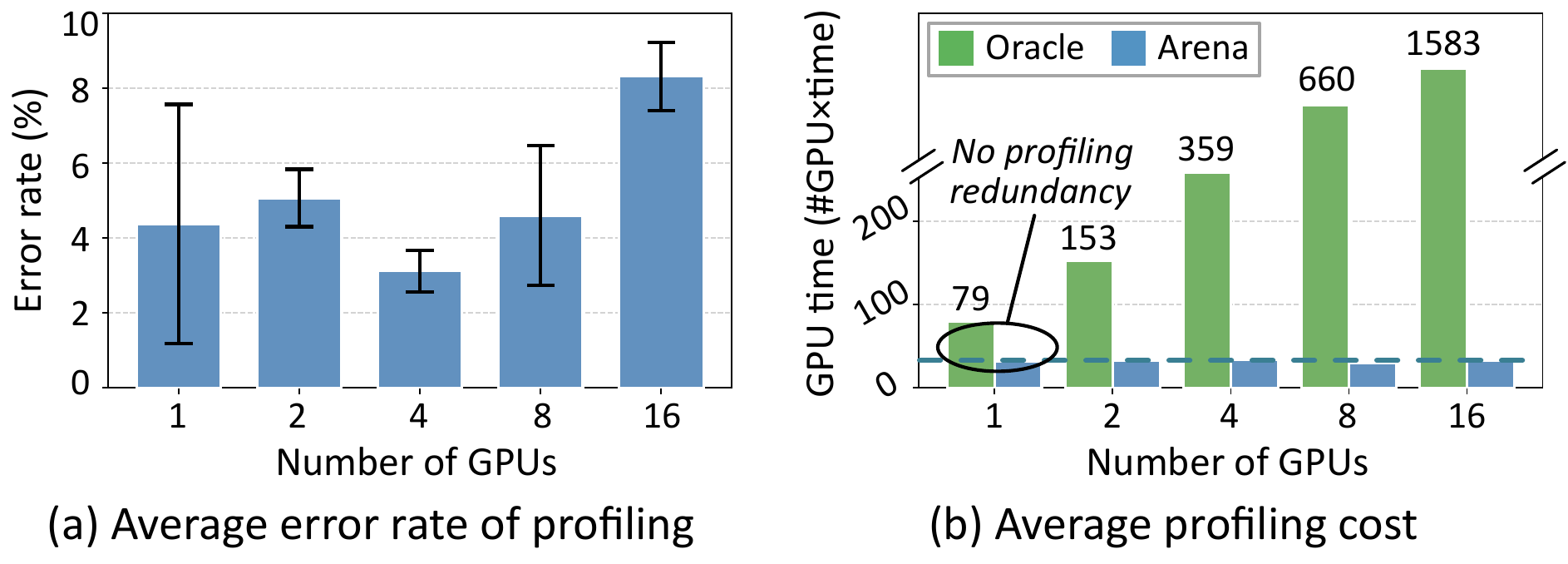}
\vspace{-2mm}
\caption{Performance of disaggregated profiling. (a) Each bar shows the average error across diverse models and hardware. (b) \texttt{Oracle} represents directly executing the model.}
\label{fig:profiler_exp}
\vspace{-1mm}
\end{figure}

Figure~\ref{fig:profiler_exp} evaluates \sysname{} profiler in terms of error rate (i.e., the gap between profiled and directly measured latencies) and profiling cost reduction.
Through precise operator profiling and end-to-end modeling, \sysname{} achieves average error rates of $4.4\%$, $5.1\%$, $3.1\%$, $4.6\%$, and $8.3\%$ for $1$, $2$, $4$, $8$, and $16$ GPU cases, sufficient for job scheduling (Figure~\ref{fig:profiler_exp}(a)).
Moreover, \sysname{} reduces the GPU time (i.e., elapsed time $\times$ occupied GPU count~\cite{mahajan2020themis}) by $18.1 \times$ on average and $2.55 \times$ at least, as compared to direct measurement.
This stems from eliminating redundant computations and device stalls, while avoiding online communication profiling that remains time-consuming under high traffic or limited bandwidth.




\subsection{Deadline-Aware Scheduling}\label{sec:generality_exp}

We also evaluate the generality of \sysname{} for other objectives by enabling deadline awareness and comparing with ElasticFlow~\cite{elasticflow}.
\sysname{} ensures strict deadlines while optimizing throughput, dropping jobs that cannot meet deadlines.
Compared with ElasticFlow, \sysname{} improves \textit{deadline satisfaction ratio} ({the proportion of jobs satisfying their deadlines}) by $1.69 \times$ and reduces JCT by $26.1\%$. 
Moreover, \sysname{} achieves $1.73 \times$ higher average throughput and $1.96 \times$ higher peak throughput.
This is attributed to the AP-aware scheduling of \sysname{}, which enables efficient resource allocation with more jobs being admitted and efficiently executed.

\subsection{Ablation Studies}\label{sec:ablation}

\begin{figure}
\centering
\includegraphics[width=.98\linewidth]{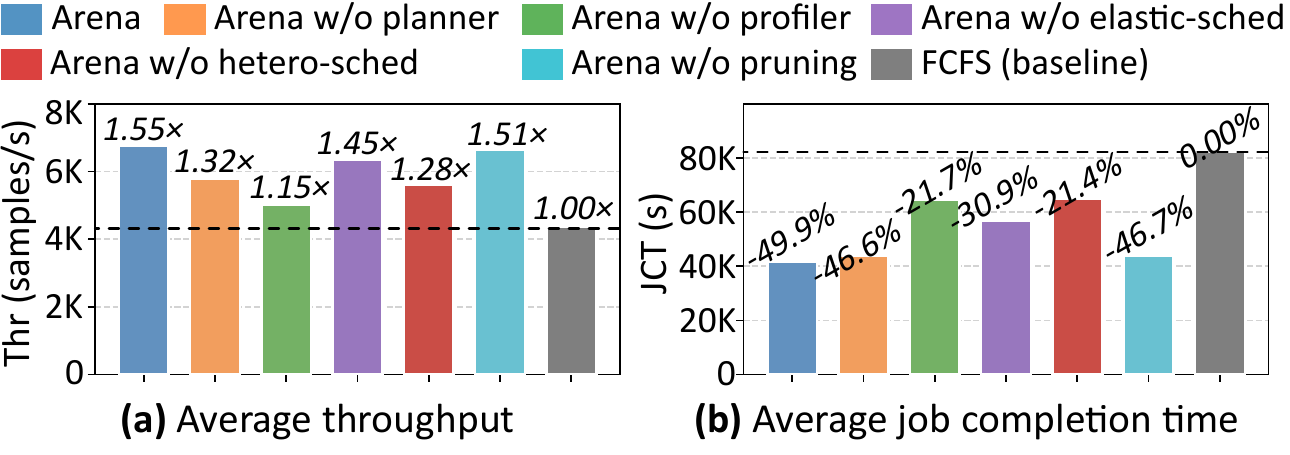}
\vspace{-2mm}
\caption{Performance breakdown of \sysname{}.}
\label{fig:ablation}
\vspace{-2mm}
\end{figure}

\paragraph{Performance Breakdown.}
To understand the sources of the benefits, we evaluate \sysname{} by sequentially disabling each component (Figure~\ref{fig:ablation}).
As the two key components of \sysname{}, the profiler and planner have notable impact on cluster throughput. 
Disabling the profiler results in $25.8\%$ less throughput and $56.3\%$ higher JCT due to aggravated resource contention between to-profile and in-flight jobs (\S\ref{sec:strawman}).
Without planner, the scheduler employs inaccurate performance data when assuming jobs are executed with DP, resulting in $14.8\%$ less throughput and $6.59\%$ higher JCT.

Another dominant component is the heterogeneity-aware scheduling, as \texttt{w/o hetero-sched} reduces cluster throughput by $17.4\%$ and increases JCT by $56.9\%$.
This impact surpasses \texttt{w/o elastic-sched} due to the high heterogeneity of the simulated testbed, which explains why Gavel outperforms ElasticFlow in simulation (4 GPU types) but falls short in real-world testbeds (2 GPU types). 
Moreover, \texttt{w/o pruning} exhibits limited performance decrease due to the limited per-job rescheduling events (2.29 in \S\ref{sec:sim}). 

\begin{figure}
\centering
\includegraphics[width=.98\linewidth]{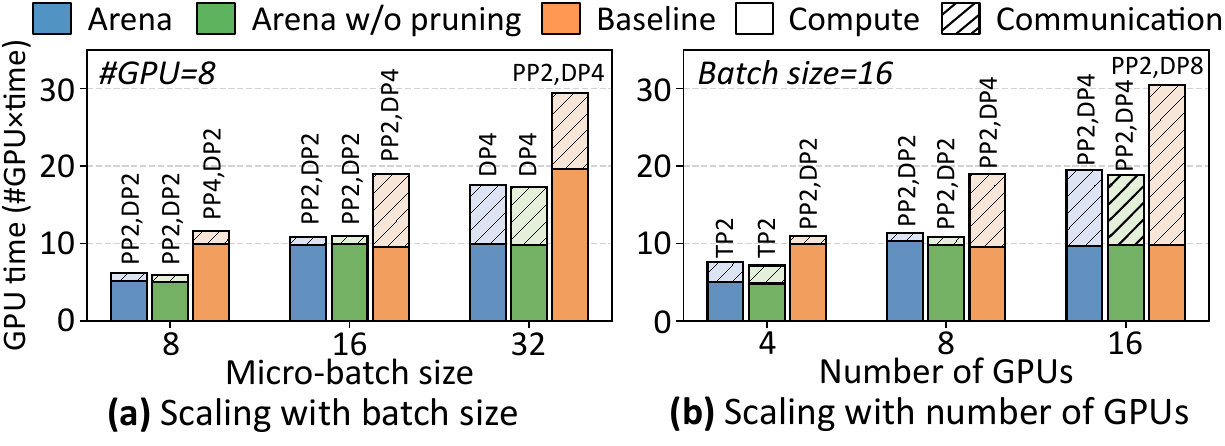}
\vspace{-2mm}
\caption{Training time breakdown of GPT-2.6B with A40 GPUs.
The searched optimal plans are marked above bars.
}
\label{fig:kernel}
\vspace{-2mm}
\end{figure}

\paragraph{Job-Level Breakdown.}
To delve into the performance gains at the job level, we breakdown iteration time into latencies of compute (e.g., GEMM) and communication kernels (e.g., all-reduce) to assess \sysname{}, \sysname{} w/o pruning, and Sia.
Givan that Sia overestimates job resources (\S\ref{sec:contradiction}), for ablation, we statically assume $2\times$ more GPUs allocated by it.
As shown in Figure~\ref{fig:kernel}, \sysname{} generates the same optimal plans with full-space AP (<$5\%$ performance gap), aligned with results in \S\ref{sec:plan_exp}.
Compared to Sia, \sysname{} reduces $41.2\%$ GPU time on average by efficiently parallelizing on less GPUs and alleviating diminishing returns in resource scaling.
For example, increasing DP has negligible impact on compute GPU time, yet largely increases communication GPU time by up to $9.15\times$ due to cross-node all-reduce operations.
Consequently, the baseline allocates more GPUs for the job but fails to proportionately improve its training efficiency.

\begin{figure}
\centering
\includegraphics[width=.98\linewidth]{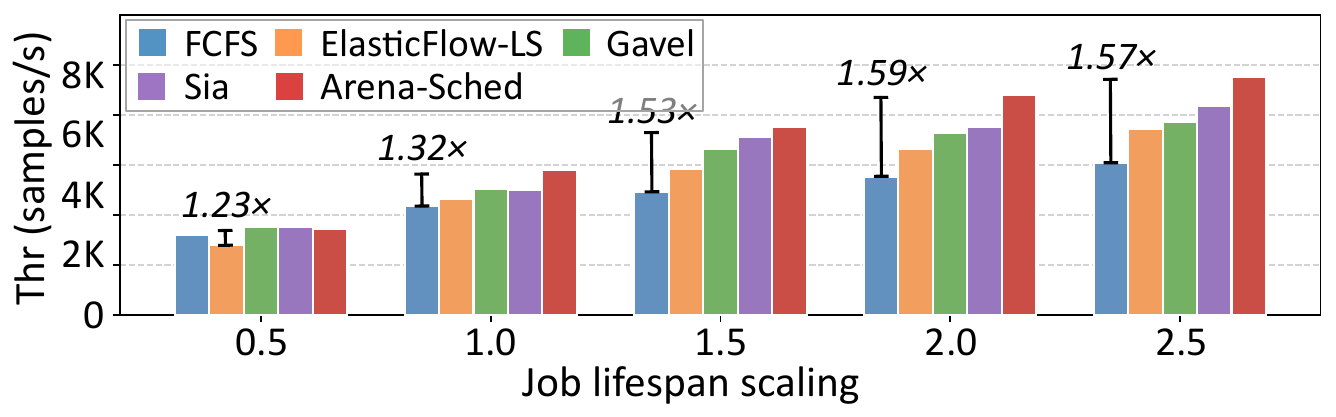}
\vspace{-2mm}
\caption{Effectiveness of \sysname{} scheduler (\sysname{}-Sched) over job lifespan scaling, other components are disabled.}
\label{fig:dync_sched}
\vspace{-2mm}
\end{figure}

\paragraph{Effectiveness of Dynamic Scheduling.}
We study the effectiveness of \sysname{} scheduler by disabling other components and scheduling based on job DP performance (aligned with baselines).
We scale job lifespan (iteration count) of the trace in \S\ref{sec:sim} for evaluation under varying workloads (Figure~\ref{fig:dync_sched}).
As job lifespan increases, the benefits of \sysname{} become more significant with up to $1.59\times$ higher average throughput, as it efficiently schedules jobs via priority-based launching, 2D scaling, and the event-driven, throughput maximization policy.
With sparse jobs (sufficient idle resources), however, the benefit is limited, as the multi-level job queues of \sysname{} fall back to FCFS and no job down-scaling is needed.


\paragraph{Homogeneous Scheduling Study.}
To illustrate the robustness of the scheduling-parallelism co-design, we further assess \sysname{} using a homogeneous real-world testbed with 128 H100 GPUs in \textit{Cluster-B}.
Evaluated with the same trace in \S\ref{sec:physical_exp} (all jobs are reassigned with only H100 GPUs), \sysname{} improves average cluster throughput by $2.36\times$ (FCFS), $1.48\times$ (ElasticFlow-LS), $2.04\times$ (Gavel), and $1.21\times$ (Sia).
These results demonstrate the benefits of \sysname{} gained from AP-aware scheduling, regardless of underlying hardware setups.


\subsection{Sensitivity Analysis}\label{sec:search_depth}

\paragraph{Impact of Parameters.}
We evaluate the impact of two parameters.
A smaller number of priority queues $P$ improves managing efficiency and reduces starvation, while a larger $P$ enhances fairness for high-priority, resource-intensive jobs.
In practice, we set $P = 3$ to reconcile the above tradeoff.
For search depth $D$ in iterative scaling, increasing $D$ from 1 to 3 raises per-job overhead from $0.88$ to $5.98$ seconds (asynchronous and non-negligible to LM training), while reducing JCT by $14.6\%$ and improving throughput by $1.03\%$.
We range $D$ between 2 and 5 in practice, with the optimal value depending on workloads and available resources. 


\paragraph{Overhead Analysis.}
We analyze three system overheads:
(\romannumeral1) \textbf{Job profiling overhead}: With multiple GPUs, the overall complexity is $O(N^2M) / O(MN) = O(N)$, further reducible via skipping repeated operators across grids and limiting grid count to $O(\log{N})$. 
Notably, industrial practice prevents overly large $N$ by assigning dedicated GPUs to hyper-scale jobs rather than dynamic scheduling.
The profiling overhead remains <20 minutes (8.5 minutes with $N=16$ and $M=4$ in evaluation~\cite{elasticflow,lucid}). 
(\romannumeral2) \textbf{Job rescheduling overhead} (non-blocking to other jobs) includes space-pruned AP search (1-2 minutes) and checkpoint-resume (<5 minutes~\cite{bytecheckpoint}), negligible as the average rescheduling count is $2.29$ during job lifetime (hours or days).
(\romannumeral3) \textbf{Offline profiling overhead} (intra-node communication primitives, one-shot before the cluster comes online) is about 3.5 hours (4-GPU node), negligible to cluster operating lifetime (weeks or months).


\paragraph{Impact on Model Convergence.}
\sysname{} ensures consistent convergence in dynamic scaling via~\cite{tenplex,gandiva,elasticflow,easyscale}: 
(\romannumeral1) fixed hyperparameters (e.g., global batch size) when adjusting DP degrees or GPU resources~\cite{elasticflow};
(\romannumeral2) streaming dataset loading that avoids re-training samples after resharding~\cite{bytecheckpoint}. 
To illustrate this, we employ \sysname{} to train a GPT-2.6B with \textit{Wikipedia} dataset~\cite{wiki_dataset} and shift the DP degree from 1 to 4 (with PP degree of 2). 
The loss curves exhibit highly similarity with the average mean-square deviation of 0.06.




\section{Discussion and Future Work}\label{sec:discuss}

\paragraph{Intra-Job Heterogeneity.}
As elaborated in \S\ref{sec:sched}, \sysname{} scheduler dynamically switches the GPU type for each job while allocating homogeneous GPUs to ensure training efficiency and manageability.
With the advent of heterogeneity-aware training frameworks (e.g., Sailor~\cite{sailor}, Metis~\cite{metis}) that parallelize large models on heterogeneous GPUs, \sysname{} remains extensible to integrate them as its training backend.
The key modifications include extending the operator load definition and GPU assignment strategy by quantifying the compute capability of different GPU types, and profiling the computation operators for each pipeline stage on a single GPU with the assigned GPU type.
We leave the exploration of integrating intra-job heterogeneity as future work.

\paragraph{Extensibility to Training Frameworks.}
\sysname{} is extensible to integrate other training frameworks (e.g., Megatron-LM~\cite{megatron}, PyTorch FSDP~\cite{pytorch-fsdp}) as the AP runtime backend.
Taking Megatron-LM as an example, several modifications are necessary since it lacks native support for adaptive parallelism and compilation.
First, it is necessary to implement APIs for flexible pipeline stage construction, per-stage GPU assignment, and stage-wise parallelism selection.
Second, the framework requires a compiler based on TorchFX~\cite{torch-fx} to obtain the intermediate representation (IR) graph of Megatron-LM models for operator disaggregation.
Lastly, the profiling of computation operators can be performed offline, given that the PyTorch dispatch mechanism guarantees consistent kernel selection for identical input shapes, data types, and hardware~\cite{torch-dispatcher}.
We have developed such a framework as a PyTorch-native training backend, which is open-source together with the JAX implementation of the \sysname{} backend.

\section{Related Work}\label{sec:related}

\paragraph{Scheduling Training Jobs.}
Cluster schedulers for deep-learning (DL) training jobs have surged recently~\cite{sia,gandiva,tiresias,hived,lucid,pai,mahajan2020themis, gandiva_fair,elasticflow,gavel,easyscale,siloD}.
ElasticFlow~\cite{elasticflow} elastically scales the number of homogeneous GPUs for in-flight jobs to boost cluster throughput while satisfying job deadlines.
Gavel~\cite{gavel} dynamically switches the type of allocated GPUs for jobs to alleviate imbalanced loads across heterogeneous GPU regions.
Sia~\cite{sia} and EasyScale~\cite{easyscale} jointly optimize both the number and type of GPUs for more flexible resource allocation.
They rely on profiling or estimation for static parallelism (primarily data parallelism), disregarding the performance of adaptive parallelism and thus failing in scheduling large models with complex parallelism strategies.

\paragraph{Parallelizing Large Models.}
Researchers have widely studied the parallelism optimization of large models~\cite{megatron,gpipe,pipedream,alpa,unity,tessel,dynapipe,metis,whale}. 
Megatron-LM~\cite{megatron} shards model parameters across multiple GPUs to reduce per-GPU memory footprint.
GPipe~\cite{gpipe} and PipeDream~\cite{pipedream} explore efficient pipeline schedules to enhance overall training throughput.
Alpa~\cite{alpa}, Unity~\cite{unity}, and Aceso~\cite{aceso} automatically search the optimal hybrid parallelism strategy on fixed resources.
While the training backend of \sysname{} is built on Alpa, its core components are compatible to various training frameworks to provide efficient and precise job performance acquisition for the co-design of scheduling and parallelization.

\paragraph{Modeling Job Performance.}
Some works~\cite{habitat,daydream,madmax,fastermoe,paleo,neusight,centimani,PCS} have studied modeling the end-to-end performance of DL jobs.
NeuSight~\cite{neusight} and Habitat~\cite{habitat} conduct blackbox prediction for kernel-level performance.
Centimani~\cite{centimani} and PCS~\cite{PCS} offline profile all operators across multiple GPUs for online modeling.
MAD-Max~\cite{madmax} analytically models the latency of communication and computation operators.
They focus on zero-GPU prediction or multi-GPU measurement under static parallelism, rather than dynamic (fragmented) resources and adaptive parallelism.

\section{Conclusion}\label{sec:conclusion}

We present \sysname{}, a co-designed training system to dynamically schedule and efficiently execute large models with adaptive parallelism in GPU clusters.
\sysname{} unifies low-cost profiling and AP-tailored estimation to efficiently navigate the scheduling-parallelism optimization space.
\sysname{} conducts dynamic AP-aware scheduling and executes jobs with space-pruned AP.
\sysname{} improves cluster throughput by $1.60\times$ and reduces JCT by $49.3\%$ as compared to baselines.

\section*{Acknowledgments}

We would like to thank the anonymous reviewers and our shepherd, Chuan Wu, for their valuable feedback. 
This work is partially sponsored by the National Key Research and Development Program of China (2024YFB4505703), National Natural Science Foundation of China (62232011), and Natural Science Foundation of
Shanghai Municipality (24ZR1430500). 
Quan Chen is the corresponding author of this paper.

\bibliographystyle{ACM-Reference-Format}
\bibliography{reference}

\clearpage
\appendix

\section{Artifact Appendix}\label{appendix-a}

\subsection{Abstract}\label{append:abstract}

\sysname{} is a large model training system to dynamically schedule and efficiently execute large models with adaptive parallelism in GPU clusters.
This artifact includes the source codes of the \sysname{} prototype and the benchmarking instructions to evaluate its functionality and reproduce its major experimental results.

\subsection{Description \& Requirements}\label{append:require}

\subsubsection{How to access.}\label{append:how-to-access}
The Arena system is available at: GitHub \url{https://github.com/sjtu-epcc/arena/tree/ae-eurosys#} and Zenodo \url{https://zenodo.org/records/18870526}. 
More detailed instructions are provided in the repository \texttt{README}.

\subsubsection{Hardware dependencies.}\label{append:hardware}
The artifact requires a Linux system equipped with at least 192 GB of system memory, 256 GB of available disk storage, and 4 NVIDIA A40 GPUs (48GB, connected via PCIe or NVLink). 
Notably, given that the full-fleet evaluation involves tens to hundreds of GPUs, for reproducibility, the artifact mainly uses 4 NVIDIA A40 GPUs unless specified.

\subsubsection{Software dependencies.}\label{append:software}
The artifact requires Conda for package management. The software stack includes CUDA 11.8 and Python3.8. All software dependencies are automatically installed in our provided Docker image.

\subsection{Set-up.}\label{append:setup}

Users should clone the repository of \sysname{} system and build from the Docker image.

\begin{lstlisting}
# Git clone
git clone --recursive -b ae-eurosys https://github.com/sjtu-epcc/arena.git
cd arena
git checkout ae-eurosys
# Docker build, run, and backend installation
cd runtime
docker build -t arena/arena:ae-eurosys -f ./profile_cu118.Dockerfile .
docker run --runtime=nvidia -it --rm --gpus all --shm-size 64g --network=host --privileged --volume [USER_DIR]/.cache:/root/.cache --env NVIDIA_DISABLE_REQUIRE=1 --name arena arena/arena:ae-eurosys
conda activate alpa         # Alpa env
bash jaxpr/cpp/install.sh   # Build cpp backend
\end{lstlisting}

\subsection{Evaluation Workflow}\label{append:evaluation}

\subsubsection{Major claims.}\label{append:claims}

The major claims of \sysname{} system for artifact evaluation include:

\begin{itemize}
    \item (\textbf{C\#1}) The disaggregated profiler of \sysname{} achieves average error
rates of $4.4\%$, $5.1\%$, $3.1\%$, $4.6\%$, and $8.3\%$ for 1, 2, 4, 8, and 16 GPU cases; Arena reduces the GPU time (i.e., elapsed time $\times$ occupied GPU count) by $18.1\times$ on average ($2.55\times$ at least), as compared to direct measurement.

    \item (\textbf{C\#2}) In the parallelism planner of \sysname{}, the best proxy plan (used for scheduling) among grids achieves average $93.4\%$ performance of the AP searched optimal plan, thus is accurate enough to achieve AP-aware cluster scheduling.

    \item (\textbf{C\#3}) With AP-aware scheduling, \sysname{} scheduler reduces average job completion time (JCT) by $81.3\%$ (FCFS), $80.5\%$ (ElasticFlow-LS), $76.6\%$ (Gavel) and $75.2\%$ (Sia), completing up to $1.45\times$ more jobs. From the cluster perspective, Arena outperforms baselines with up to $1.55\times$ higher average throughput and $1.58\times$ higher peak throughput.

\end{itemize}

\subsubsection{Disaggregated profiling experiment (\textbf{C\#1}).}
We provide the instructions to run the single-device profiler and the multi-device direct execution to evaluate the accuracy and profiling cost reduction.

Before running single-device profiling, users should offline profile the communication latency data (may take dozens of minutes or a few hours). Detailed instructions are provided in repository \texttt{README}.

Users can evaluate single-device profiling by specifying (1) model configurations (layers are uniformly clustered into stages), (2) device assignments, and (3) parallelism plans.
Taking vanilla 1F1B pipeline parallelism on 4 GPUs as an example (\textbf{full instructions in repository \texttt{README}}, ``Crius'' is the alias for ``\sysname{}'', ``cell'' for ``grid''):

\begin{lstlisting}
export ENABLE_CRIUS_PROFILER=true   # Enable arena
export CUDA_VISIBLE_DEVICES=0   # Specify a single GPU
python jaxpr/runtime_profiler.py --estimate_e2e --num_hosts 1 --num_devices_per_host 4 --devices_name 1_a40 --model_name wide_resnet --param_num 1B --batch_size 256 --parallel_degrees=4,1,1   # [pp,dp,tp]
\end{lstlisting}

To measure the end-to-end iteration time (rather than estimating it) with the specified parallel plan:

\begin{lstlisting}
export CUDA_VISIBLE_DEVICES=0,1,2,3 # Specify all
ray start --head    # Start ray cluster
export ENABLE_CRIUS_PROFILER=false  # Disable arena
python jaxpr/runtime_profiler.py --measure_with_alpa --num_hosts 1 --num_devices_per_host 4 --devices_name 1_a40 --model_name wide_resnet --param_num 1B --batch_size 256 --parallel_degrees=4,1,1  # [pp,dp,tp]
\end{lstlisting}

The evaluation results are output in the console logs. 
In our 1$\times$4 A40 node, the estimated/measured end-to-end iteration time is 15.893s/16.121s with the profiling cost of 63.475/613.93 \#GPU $\times$ time(s).

\subsubsection{Parallelism planning experiment (\textbf{C\#2}).}
We then provide the instructions to run the parallelism planner of \sysname{} to evaluate the performance of the best proxy plan (used for cluster scheduling) among all grids, compared to the AP searched optimal plan from Alpa. 

To evaluate the parallelism planner among all grids, i.e., the best proxy plan (used for scheduling this job) among all grids:

\begin{lstlisting}
export ENABLE_CRIUS_PROFILER=true   # Enable arena
export CUDA_VISIBLE_DEVICES=0   # Specify one GPU
python jaxpr/crius_cell_profile.py --num_hosts 1 --num_devices_per_host 4 --devices_name 1_a40 --model_name wide_resnet --param_num 1B --batch_size 256 --num_micro_batches 16 --cell_prof_strategy=auto
\end{lstlisting}

To measure the performance of the optimal parallelism plan searched by Alpa:

\begin{lstlisting}
export CUDA_VISIBLE_DEVICES=0,1,2,3 # Specify all
ray start --head    # Start ray cluster
export ENABLE_CRIUS_PROFILER=false  # Disable arena
python jaxpr/runtime_profiler.py --optimize_with_alpa --num_hosts 1 --num_devices_per_host 4 --devices_name 1_a40 --model_name wide_resnet --param_num 1B --batch_size 256
\end{lstlisting}

The evaluation results are output in the console logs. 
In our 1$\times$4 A40 node, the end-to-end iteration times of \sysname{} estimated and Alpa optimized are 10.445s and 10.633s, respectively.

\subsubsection{Large-scale simulated scheduling experiment (\textbf{C\#3}).}
The simulation requires offline profiling all training jobs by enumerating possible combinations of models (e.g., GPT-1.3B), hyperparameters (e.g., global batch size 256), and allocated hardware (e.g., 1$\times$4 A40 GPUs). 
Here, to avoid extensive offline profiling for artifact evaluation, we have provided our profiling data in \texttt{./database/prof\_database.pkl} that includes \sysname{}'s estimated data, data profiled via data parallelism, and Alpa's searched data. 
Notably, \textbf{running this experiment requires no GPU resources}, and users can either copy the scheduler-related codes into the container via \texttt{docker cp ./ arena:/app/} or directly execute the simulator on the host (need to manually install minor related dependencies).

To run large-scale simulated scheduling with 1,280 GPUs and Philly trace, use the following instructions (\texttt{[POLICY]} includes \texttt{fcfs}, \texttt{elasticflow-l}, \texttt{gavel}, and \texttt{sia}):

\begin{lstlisting}
# For Arena
python simulator.py --policy=crius --trace_type=philly --sched_with_opt --max_sched_round=2000 --enable_alpa --result_dir=./plot
# For other baselines
python simulator.py --policy=[POLICY] --trace_type=philly --max_sched_round=2000 --enable_alpa --result_dir=./plot
\end{lstlisting}

We also provide scripts to visualize the results (\texttt{[METRIC]} includes \texttt{thr}, \texttt{jct}, and \texttt{queuing\_time}):

\begin{lstlisting}
python simulator.py --visual --visualized_metric=[METRIC] --result_dir=./plot --out_dir=./figures --trace_type=philly
\end{lstlisting}

For throughput metric, users can both inspect average/maximum values in console logs and visualized curves (Figure~\ref{fig:overall_thr}) in \texttt{./figures/cluster\_thr.pdf}.
For JCT, number of finished jobs, and queuing time metrics, users can inspect results in console logs.

\end{document}